\documentstyle{mn}
\title[SuperCOSMOS Sky Survey II: Image detection \& parameterisation etc.]
{The SuperCOSMOS Sky Survey. Paper II: \\
Image detection, parameterisation, classification and\\
photometry}
\author[N.C.\ Hambly et al.]{N.C.\ Hambly$^{\rm 1}$,  
M.J.\ Irwin$^{\rm 2}$ and H.T.\ MacGillivray$^{\rm 1}$\\
$^1$Wide Field Astronomy Unit, Institute for Astronomy, University of Edinburgh, Blackford Hill, Edinburgh, EH9~3HJ\\
$^2$Cambridge Astronomy Survey Unit, Institute of Astronomy, Madingley Road,
Cambridge, CB3~0HA\\
}

\date{Accepted ---. 
      Received ---;
      in original form ---}

\begin{document}

\maketitle

\begin{abstract}
In this, the second in a series of three papers concerning the SuperCOSMOS
Sky Survey, we describe the methods for image detection, parameterisation,
classification and photometry. We demonstrate the internal and external accuracy
of our object parameters. Using examples from the first release of data, the
South Galactic Cap survey, we show that our image detection completeness is
close to 100\% to within $\sim1.5$~mag of the nominal plate limits. We show
that for the B$_{\rm J}$ survey data, the image classification is 
{\em externally} $>99$\% reliable to B$_{\rm J}\sim19.5$. Internally, the image
classification is reliable at a level of $>90$\% to B$_{\rm J}\sim21$, R~$\sim19$.
The photometric accuracy of our data is typically $\sim0.3$~mag with respect to
external data for m~$>15$. Internally, the relative photometric accuracy in
restricted position and magnitude ranges can be as accurate as $\sim5$\% for
well exposed stellar images. Colours are externally accurate to 
$\sigma_{\rm B-R,R-I}\sim0.07$ at B$_{\rm J}\sim16.5$ rising to
$\sigma_{\rm B-R,R-I}\sim0.16$ at B$_{\rm J}\sim20$.

\end{abstract}

\section{Introduction}
\label{intro}

In Paper~{\sc I} of this series (Hambly et al.~2001a) we describe the SuperCOSMOS
Sky Survey programme (hereafter SSS). This project is to scan the 
multi--colour/multi--epoch Schmidt photographic atlas material to produce a
digitised survey of the sky in three colours (BRI), one colour (R) at two epochs.
The ultimate aim of the project is to cover the entire sky; the 
first release of data from the programme was the South Galactic Cap (hereafter
SGC) survey. The SGC survey covers $\sim5000$ square degrees of the southern
sky at Galactic latitudes $|b|>60^{\circ}$. Paper~{\sc III} in this series
describes the derivation of the astrometric parameters for the SSS
(Hambly et al.~2001b). In this,
the second paper of the series, we describe in some detail the image
detection, parameterisation and classification procedures for the programme.
We also describe the techniques for photometric calibration. 

Paper~{\sc I} is intended as a User Guide for the survey data. It describes the
database organisation and demonstrates specific examples of the use of the data.
Papers~{\sc II} and~{\sc III} provide technical details concerning the
derivation of object catalogue parameters available from the survey database,
and also demonstrate the precision of these with respect to external data from
other sources.  In these papers, we demonstrate examples and make comparisons
using data from the SGC survey. All the results, however, are generally 
applicable to the SSS data as a whole at Galactic latitudes 
$|b|\geq30^{\circ}$. At lower latitudes, image crowding will of course degrade
astrometric and photometric performance.

The plate material used in the SSS is detailed in Paper~{\sc I}, and 
consists of sky--limited Schmidt photographic glass plate and film originals,
or glass copies of glass originals, taken with the UK, ESO and Palomar 
Oschin Schmidt
Telescopes (for more details see Morgan et al.~1992 and references therein).
Hereafter, the SERC--J/EJ survey will be refered to simply as the J~survey;
the AAO--R/SERC--ER as the R~survey and the SERC--I as the I~survey.

\section{Methodology}
\label{methods}

\subsection{Pixel analysis}

\subsubsection{Digitisation}

The SuperCOSMOS machine digitises the scanned photograph at 10$\mu$m resolution
(0.67~arcsec)
in 15--bit grey levels (further details of the hardware are given in Hambly
et al.~1998). Pixel values are in transmission units and are related to
incident intensity via the usual transmission ($T$), density ($D$) and intensity
($I$) relationships on the the log--linear part of the photographic 
characteristic curve:
\begin{equation}
\gamma\log_{10}I\propto D \propto\log_{10}(1/T), 
\end{equation}
where $\gamma$ is the gradient of the response in the $D-\log_{10}I$ plane.
Each photograph is digitised in $\sim2$~hr, producing $\sim2$~Gbyte of data.
The pixel map is `blocked down' in $64\times64$ pixel cells during scanning
to produce a raw median sky map; all subsequent processing is then carried out
offline.

\subsubsection{Sky estimation}

Object detection in the raw pixel map requires an estimate of the sky value at
every point within the two--dimensional image. The raw median values of the sky
are first filtered using a weighted median filter (Brownrigg~1984) of the form
\[\begin{array}{ccc}
1 & 1 & 1 \\
1 & 3 & 1 \\
1 & 1 & 1 \\
\end{array} \]
applied over filtering scales of 3,2 and~1 blocks (e.g.~Stobie~1986). This
results in
variations in scale size greater than $\sim2$~mm (or $\sim2$~arcmin) being
retained as sky variations in the background array. This is obviously a 
compromise between detection of low surface brightness objects at this scale
or greater and accurate image parameterisation in areas of real sky variation.
Experiments with differing scale lengths showed that the choice of raw blocking
of 0.64~mm squares coupled with iterative filtering scale sizes of~3,2 and~1 
(e.g.~Stobie~1986) is optimal in terms of minimising spurious detections
due to, for example, low level scattered light around bright stars (see also
MacGillivray \& Stobie~1984).

\subsubsection{Transmission to intensity calibration}

Most modern Schmidt photographs have one or more intensity calibration 
`step--wedges' exposed on their sides or corners (see, for example, 
Tritton~1983). In principle, it is possible to measure the transmission of the
various intensity spots and calibrate measured transmission into
incident relative intensity. In practice, however, there are a number of reasons
why even rough step--wedge calibration is difficult to achieve. Variations in
emulsion sensitivity across the plate and non--uniform illumination of the wedge
spots themselves can cause large errors. Also, the fact that SuperCOSMOS 
scans by illumination of a strip of emulsion means that when imaging dense
spots, the light levels recorded are badly affected by diffracted light 
originating from relatively clear emulsion surrounding the spot. Measurements
of the 16--spot KPNO step wedges on modern UKSTU plates using a Macbeth diffuse
densitometer show values of characteristic curve gradient $\gamma\sim2\pm10$\%
(indeed, such a value is a criterion for a survey--grade plate and inclusion in
the final UKSTU photographic sky atlases). Hence, we use an {\em assumed}
transmission--to--intensity calibration with slope corresponding to
$\gamma\approx2$ (older glass copies of the POSS--I `E' survey
are analysed using a different value due to their poorer contrast). This
procedure is vindicated not least by the demonstration, in Section~\ref{grads},
of measurements of the gradients of the instrumental magnitude scale with
respect to CCD data, where we show that the resulting scales have close to unit
gradient with scatter at around the 10\% level.

\subsubsection{Thresholding and image parameterisation}
\label{thresholding}

Image detection requires thresholding above a local sky estimate. The SSS
pixel data are thresholded at a constant percentage cut above local
sky. A sky noise parameter $\sigma_{{\rm SKY}_0}$ is determined from a small
($\sim10\times10$~arcmin) region of image data near the centre of each plate.
The sky background value I$_{{\rm SKY}_0}$ 
in this region is determined as described in Beard,
MacGillivray \& Thanisch~(1990) while the $\sigma_{{\rm SKY}_0}$ estimator uses
the relationship $\sigma_{{\rm SKY}_0}=1.48\times{\rm MAD}_{{\rm SKY}_0}$, where
MAD$_{{\rm SKY}_0}$ is the median of absolute deviations of the pixel values
about the sky estimate (the factor 1.48 is the scaling needed to convert the MAD
estimate for a Gaussian distribution to the equivalent Gaussian sigma).
This yields a `percentage cut' value PCUT 
($=100\times2.3\sigma_{{\rm SKY}_0}/I_{{\rm SKY}_0}$) 
such that at any part of the image
\begin{equation}
I_{\rm THRESH}=\frac{(100+{\rm PCUT})}{100}\times I_{\rm SKY}, 
\label{threshold}
\end{equation}
where $I_{\rm THRESH}$ is the threshold intensity, $I_{\rm SKY}$ is the local
sky estimate based on bilinear interpolation of the filtered sky map, and
PCUT is typically 7\% for
survey plates. Note that this single--parameter multiplicative thresholding is
designed to cope with the fact that the sky level (and therefore the noise
level) changes over the Schmidt field due to vignetting, and a zeropoint sky
intensity value is chosen for the transmission--to--intensity calibration
such that the threshold, in terms of the number of $\sigma$ above sky, remains
constant regardless of sky variations.

Once the thresholded pixel set has been determined for a plate, pixel 
connectivity can be analysed and image parameterisation can take place. The
details of the COSMOS pixel analyser are contained in MacGillivray \& 
Stobie~(1984), Stobie~(1986) and Beard et al.~(1990) and will not be repeated
here; however we describe some details that are relevant to SuperCOSMOS survey
data. Many of the techniques used are similar to those developed for the APM
machine (e.g.~Irwin~1985; Draper \& Eaton~1999).

For image detection, 8--fold nearest--neighbour pixel connectivity is used along
with a criterion that images must contain at least~8 connected pixels (the 
so--called `area cut'). Rethresholding at 16 levels defined as in Beard et 
al.~(1990) Equation~2, with a `deblend parameter' of D$=1.05$ detects
multiple objects via fragmentation; in addition rethresholding at 8 levels
defined by
\begin{equation}
I_{{\rm AP}n} = I_{\rm THRESH} + 2^{n-2}\times\sigma_{{\rm SKY}_0}, n=1,2,...8 
\label{areal}
\end{equation}
is used to supply a set of 8 `areal profiles' at intensity levels of 
$0.5\sigma$, $1\sigma$, $2\sigma$, $4\sigma$ \ldots above the zeroth areal
profile level $I_{\rm THRESH}$. These provide a set of measurements of the areal
extent (in pixels) of each image, and can be used for morphological 
classification (see Section~\ref{class}). Note that these areal profile levels
are different to those originally used for the COSMOS machine (MacGillivray
\& Stobie~1984). This new definition provides optimal levels for stellar
profile analysis (see later).

Object parameterisation proceeds for each detected object as described in
Stobie~(1986), with both unit--weighted and intensity--weighted parameters
being calculated and
recorded. The object catalogue produced for each plate consists of an
unformatted binary file (the Image Analysis Mode, or IAM, file) of record
length 128~bytes consisting of 32 4--byte parameters per object; depending on
Galactic latitude and plate depth, this individual plate catalogue can have
anything from~$10^5$ to~$10^7$ entries. Table~\ref{iam} details the full
32 parameter set available for each image. Note that parameters 1,2,19,20,31
and~32 and some bits in parameter~30 are derived at a second stage of
processing (known colloquially as `post processing' -- see Section~\ref{pp})
\begin{table*}
\begin{center}
\begin{tabular}{lllll}
No. & Name & Description & Units & Comments \\
\\
 1 & RA       & Celestial Right Ascension & $10^{-8}$~radians & see Section~\ref{astrom} \\
 2 & DEC      & Celestial Declination     & $10^{-8}$~radians & see Section~\ref{astrom} \\
 3 & XMIN     & Left extent               & 0.01 micron & \\
 4 & XMAX     & Right extent              & 0.01 micron & \\
 5 & YMIN     & Bottom extent             & 0.01 micron & \\
 6 & YMAX     & Top extent                & 0.01 micron & \\
 7 & AREA     & Total area                & Pixels & \\
 8 & IPEAK    & Peak intensity            & Intensity & $I_{\rm PEAK}$ \\
 9 & COSMAG   & Isophotal magnitude       & Millimag & Equation~\ref{cosmag} \\
10 & ISKY     & Sky intensity at (XCEN\_I,YCEN\_I) & Intensity & $I_{\rm SKY}$,
e.g.~Equation~\ref{threshold} \\
11 & XCEN\_I  & Intensity weighted X centroid & 0.01 micron & \\
12 & YCEN\_I  & Intensity weighted Y centroid & 0.01 micron & \\
13 & A\_U     & Unweighted semi-major axis    & 0.01 micron & \\
14 & B\_U     & Unweighted semi-minor axis    & 0.01 micron & \\
15 & THETA\_U & Unweighted orientation        & Degrees & \\
16 & A\_I     & Weighted semi-major axis      & 0.01 micron & \\
17 & B\_I     & Weighted semi-minor axis      & 0.01 micron & \\
18 & THETA\_I & Weighted orientation          & Degrees & \\
19 & CLASS    & Classification~flag           & & see Section~\ref{class} \\
20 & P\_A     & Celestial position angle      & Degrees & see Section~\ref{astrom} \\
21 & AP(1)    & Area above areal profile level 1 & Pixels & Equation~\ref{areal} \\
22 & AP(2)    & Area above areal profile level 2 & Pixels & \\
23 & AP(3)    & Area above areal profile level 3 & Pixels & \\
24 & AP(4)    & Area above areal profile level 4 & Pixels & \\
25 & AP(5)    & Area above areal profile level 5 & Pixels & \\
26 & AP(6)    & Area above areal profile level 6 & Pixels & \\
27 & AP(7)    & Area above areal profile level 7 & Pixels & \\
28 & AP(8)    & Area above areal profile level 8 & Pixels & \\
29 & BLEND    & Deblending flag                  & & see text \\
30 & QUALITY  & Quality flag                     & & see text \\
31 & N(0,1)   & Profile classification statistic & $\sigma\times1000$ & see Section~\ref{class} \\
32 & PRFMAG   & Profile magnitude                & Millimag & see Section~\ref{class} \\
\end{tabular}
\caption[]{The full 32--parameter Image Analysis Mode set for each detected
image, along with units for internal storage as 4--byte integers.}
\label{iam}
\end{center}
\end{table*}
Most of the parameters in Table~\ref{iam} are self--explanatory and are
derived directly from first-- and second--order moments as described in
Stobie~(1986). However, the following gives further details of some of the more
esoteric parameters:
\begin{itemize}
\item
COSMAG (parameter~9) is calculated from the sky--subtracted intensity values
in the connected pixels comprising an image,
\begin{equation}
{\rm COSMAG}=\{-2.5\log_{10}\sum_{i=1}^{n}(I_{i}-I_{\rm SKY})\}\times10^3,
\label{cosmag}
\end{equation}
and is given in units of millimags. COSMAG is an isophotal magnitude.
\item
BLEND: each image is deblended by rethresholding at the levels defined by
Equation~2, Beard et al.~(1990);
at each level, the connected pixels are examined for
fragmentation and pixels are assigned to fragments using the algorithm
described therein. Images which have been deblended have this
indicated in the deblending flag. A code of $-n$ means the image is the parent
image of $n$ components, while a code of $+n$ means that the image is the
$n^{\rm th}$ deblended component (or child image) of a parent image. Both
parents and children are retained in the IAM datasets of each plate in the SSS.
\item 
QUALITY is an integer flag of 32 bits that are used to record situations
encountered during the analysis of the pixels comprising an image. Bits are
set in such a way that increasingly severe conditions are flagged by setting
increasingly significant bits in the quality flag. Table~\ref{qbits} gives
more information; other bits in the quality flag are either used only for
image analysis software debugging or are totally unused.
\end{itemize}
\begin{table*}
\begin{center}
\begin{tabular}{lllll}
Event & Bit & Incr- & Severity & Comments \\
      & set & ement &          & \\
      &     &       &          & \\
Orientation calculation failed & 0 & 1 & Information & Image perfectly round \\
Ellipticity calculation failed & 1 & 2 & Information & Image perfectly straight\\
Image too multiple for deblending & 2 & 4 & Warning & Image split into too many
fragments \\
Bright image & 4 & 16 & Information & Image has pixels brighter than highest areal profile level \\
             &   &    &             & (as defined by $n=8$ in Equation~\ref{areal}) \\
Large image & 6 & 64 & Warning & Image has area greater than maximum specified
for deblending \\
Possible bad image & 7 & 128 & Warning & Image is in region likely to be 
affected by step wedge or label \\
                   &   &     &         & (see Section~\ref{addbits}) \\
Image near very bright star & 10 & 1024 & Warning & Image may be spurious due to
bright star \\
                            &    &      &         & (see Section~\ref{addbits})\\
Image touches boundary & 16 & 65536 & Severe defect & Image may be partially
missing \\
\end{tabular}
\caption[ ]{IAM parameter 30 (quality flag) details.}
\label{qbits}
\end{center}
\end{table*}

\subsection{Post--processing}
\label{pp}

The post--processing of each field consists of a number of separate tasks that
either overwrite information into the individual plate catalogue files, or that
write separate files of data or coefficients. Paper~{\sc I} describes the SSS
database organisation and the system that is used
to associate individual plate files into a seamless whole. Here, we describe
the post--processing tasks that operate on each plate datafile to produce
astrometric, photometric and image quality/class information.

\subsubsection{Astrometry: parameters 1,2 and 20}
\label{astrom}

Paper~{\sc III} describes the derivation of global astrometric solutions for 
each plate. IAM parameters 11,12 and~18 (respectively $x$, $y$ and ellipse--fit
position angle with respect to the $x$ axis) are transformed into celestial 
co-ordinates and a celestial position angle which are written 
to IAM parameters 1,2 and~20 respectively. Proper motion information is also 
derived between J/R plate pairs and held in separate files, as described in 
Paper~{\sc III}.

\subsubsection{Additional quality information}
\label{addbits}

Two additional bits in the quality flag that are unused by the image analysis
software are derived at the post--processing stage. Firstly, bit~7 is used
to flag images that are in parts of the plate that are likely to be affected
by the presence of step--wedges and labels. Areas of the plate are flagged
in this way based on the number and the telescope origin of the plate. For
example, all UK Schmidt plates have a 7--step wedge in the west of their
northern edge; in addition plate numbers 1 through to 5715 have a 7--step
wedge on their eastern edge; plate numbers 5716 through to 6295 have a 16--step
KPNO wedge in the same eastern position of the 7--step wedge while all plates
numbered from 6296 have a 16--spot KPNO wedge in their south--eastern corner.
The survey plate sets used have been examined for wedges and labels, and all
images in the appropriate areas have bit~7 set to indicate possible
spurious nature or probable poorly determined parameters
in the case of real images. Because the modern
survey field centres are spaced to provide $\sim0.5^{\circ}$ overlap regions
between adjacent plates, all parts of the sky affected by a wedge or label on
one plate will be unaffected on one or other of the 
adjacent plates, so these flags are used
when creating `seamless' catalogues across plate overlap boundaries when
making a decision as to which image to include out of a set of duplicates
(see Paper~{\sc I}).

\begin{figure*}
\begin{center}
\setlength{\unitlength}{1mm}
\begin{picture}(150,90)
\includegraphics{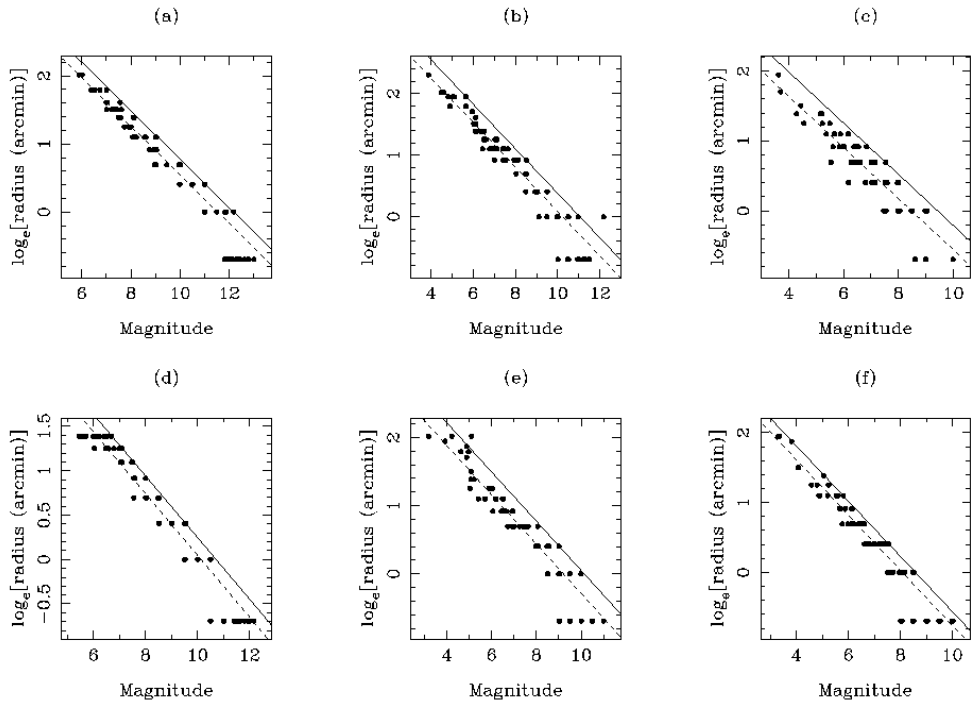}
\end{picture}
\end{center}
\caption[]{Exclusion radii as a function of magnitude for (a) SERC--J/EJ;
(b) SERC--ER/AAO--R; (c) SERC--I; (d) AAO--R on 4415 film; 
(e) ESO--R and (f) POSS--I E surveys. The dashed lines are linear least--squares
fits to the data while the solid lines are parallel to the fits but offset 
upwards to provide a conservative exclusion radius. 
Objects within the exclusion radius of a bright star have bit~10 set in
their quality flag to indicate possible spurious nature.}
\label{radfit}
\end{figure*}

Secondly, possible problems with an image due to the proximity of a very bright
star are indicated using bit~10. Tritton~(1983) describes in some detail the
origin and appearance of various image defects around bright stars on 
sky--limited UK Schmidt plates. While the background following algorithm is
tuned to minimse spurious image detections around bright stars, there are
inevitably many false detections due to, for example, diffraction spikes.
Within the SGC survey, 10 plates from each survey colour were examined by eye
to measure an exclusion radius, as a function of magnitude, around bright stars.
In Figure~\ref{radfit} we show these measurements and the log--linear fit as
a function of magnitude that is used to flag images as potentially spurious.
We have been conservative in this procedure in the sense that we have flagged 
all images that could {\em possibly} be spurious, including of course some real
images. Hence, if a relatively uncontaminated (but incomplete)
catalogue is required to be
extracted from the survey data, this quality bit can be tested. The flagging
procedure is completely automatic: given the fit as a function of magnitude and
a brightness threshold above which to apply it, a list of bright stars is
extracted from each IAM dataset and a record kept of the magnitude and radius
for flagging. These lists are recorded for each field and can be used for
estimating spatial sampling completeness when employing this flag.

\subsubsection{Image classification}
\label{class}

The SSS data are image classified using a two--stage classification
scheme originally developed for APM data (e.g.~Irwin \& McMahon 1992). The 
first classification stage makes use of conventional techniques such as surface
brightness and image shape (cf.~Heydon--Dumbleton, Collins \& 
MacGillivray~1989; MacGillivray \& Stobie~1984). The second stage uses a 
series of image parameters to trace the stellar locus in a multidimensional
parameter space and then quantify the similarity, or otherwise, of an 
individual image with respect to stellar objects of the same intensity.
Advantages of this second stage include: the generation of a `linearised' 
instrumental magnitude scale (see later); the ability to incorporate prior 
knowledge of the expected fraction of different classes of images; and a 
stellarness index in addition to the more usual discrete classification flag
(stellar/non--stellar/noise). The stellarness index quantifies how well the 
parameters of a given image match the stellar template for that intensity.
For ease of use, the index is given in terms of the equivalent normalised 
Gaussian distribution, of zero-mean and unit-variance -- ie. an N(0,1) 
statistic.

The first stage of image classification consists of generating scatter plots 
using a combination of measures of the image magnitudes, surface brightness
and peak brightness versus magnitude and
automatically locating the stellar locus in each. The stellar locii are 
automatically found by converting each scatter plot into a 128 $\times$ 128 
two-dimensional map, where each ``pixel'' repesents the number of points 
lying in that bin.  Obvious noise-like objects, for example with ellipticity
$>$0.9 or with areal profile slopes incompatible with real images, are not
included in the two-dimensional count (but are shown in the the scatter 
plots).  The location of the mode, and scatter about the modal value, are 
then determined, where possible, for each magnitude/peak slice
and used to track the stellar locus and spread in stellar locus.
For faint magntiudes where the stellar locus blends in with the remaining 
object locii, the centre-of-gravity of the generic distribution is used 
instead.

\begin{figure*}
\begin{center}
\setlength{\unitlength}{1mm}
\begin{picture}(150,220)
\includegraphics{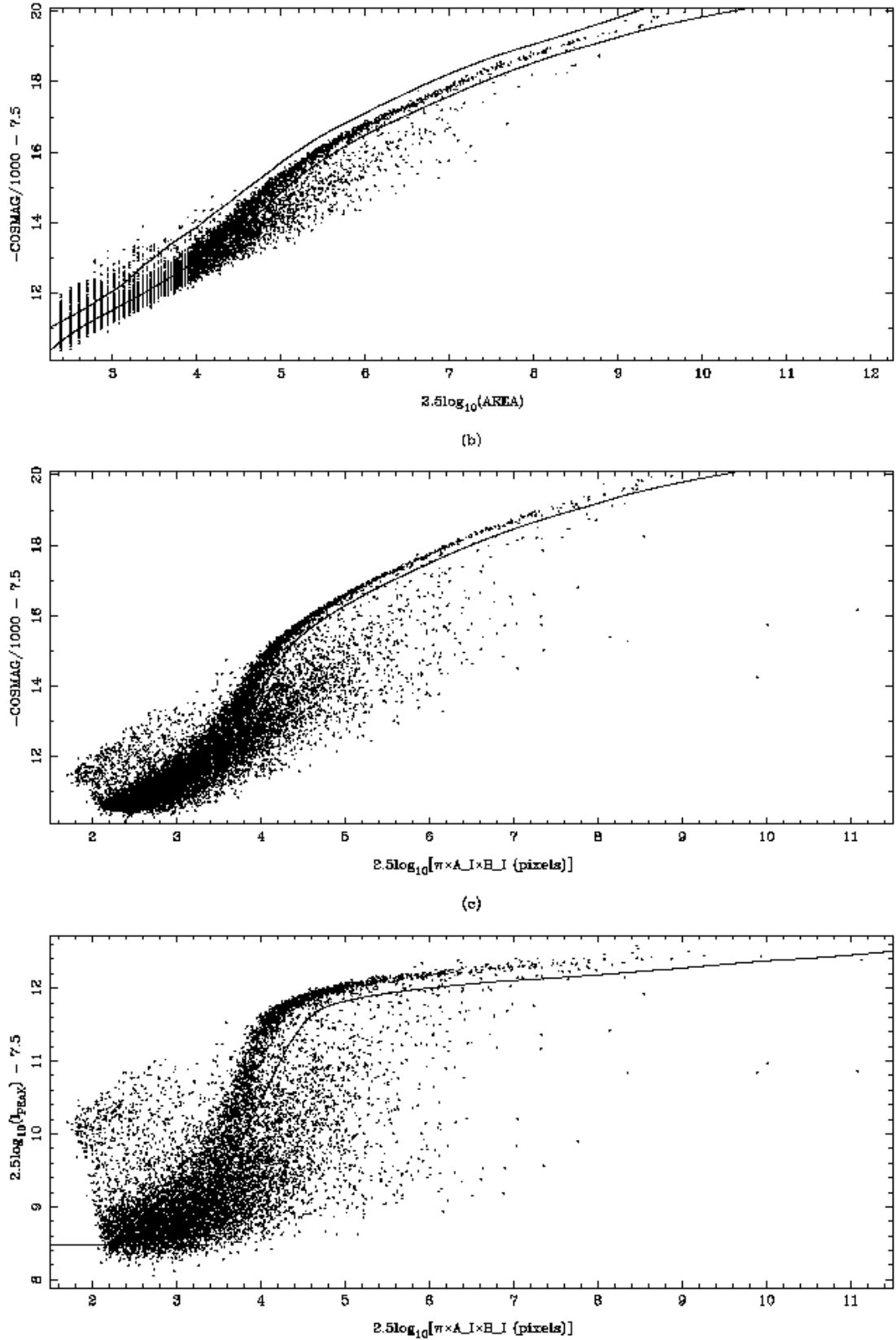}
\end{picture}
\end{center}
\caption[]{Scattergrams of image magnitude/peak
brightness versus unweighted/weighted area used at the first
stage of image classification.}
\label{stats}
\end{figure*}

Figures~\ref{stats}(a) to~(c) show typical examples of these scatter plots, 
where the solid lines mark the boundaries of the stellar/non-stellar locus
-- defined to be $2\sigma$ away from the mode, where the mode is well 
measured, or defined as the centre of gravity of the distribution 
for the faint end.  The isophotal area versus magnitude diagram is 
additionally used to define a stellar/noise image boundary which is further
used to flag images not to consider in the other two plots.
After some experimentation, a combination of the three scatter plots shown in 
Figures~\ref{stats} was found to give a reliable stage-1 classification over 
a wide range of magnitudes, Galactic location and plate/emulsion combinations.

The final stage~1 classification is based on a combination of the signed 
sigma-normalised distances from the stellar locii from these three plots 
plus the image ellipticity information.  As noted previously, at this stage 
the stellar/non--stellar boundary is set to conservatively class objects as 
stellar by defining a $2\sigma$ boundary from the median stellar locus.  
In a final step, obvious non-stellar objects in the stellar bin are rebadged
by checking for consistency of other parameters such as ellipticity and
areal profile slope.

At the completion of the first stage of image classification, each image has
IAM parameter~19 overwritten by a classification code in the range~1 to~4 
where 1=non--stellar, 2=stellar, 3=unclassifiable and 4=noise (i.e.~images 
lying above the upper locus in Figure~\ref{stats}a.)  

The provisional classification from stage~1 is then used as the primary 
input for the stage~2 classification scheme outlined earlier.  This second 
phase builds upon the classification already developed by using it as the 
basis for deriving a more general descriptor of image shape.  Providing the 
stage~1 stellar sample is a fair representation of the total stellar 
population, and only has modest contamination by non-stellar images, it is 
feasible to use it as a means to automatically track the stellar locus for 
any combination of the measured SuperCOSMOS image parameters. 

Automatically deriving both the position and spread in the stellar locus 
as a function of magnitude, for any of the image shape parameters is 
straightforward, even with modest amounts of non-stellar object contamination, 
providing robust estimation techniques are used (e.g.~Hoaglin, Mosteller 
\& Tukey 1983). For any desired image shape descriptor, the location
of the median of the population suffices to define the locus as a function
of magnitude, while the median of the absolute deviation from the median 
(MAD) provides a good measure of the scatter about this locus.  The MAD
estimate of scatter can then be converted to the equivalent Gaussian sigma
by noting, as before, 
that $\sigma_{\rm Gauss} = 1.48\times{\rm MAD}$.  By this means each of the
individual image shape descriptors can be renormalised to follow a 
zero-median, unit-variance Gaussian-like N(0,1) distribution.  

\begin{figure*}
\begin{center}
\setlength{\unitlength}{1mm}
\begin{picture}(150,90)
\includegraphics{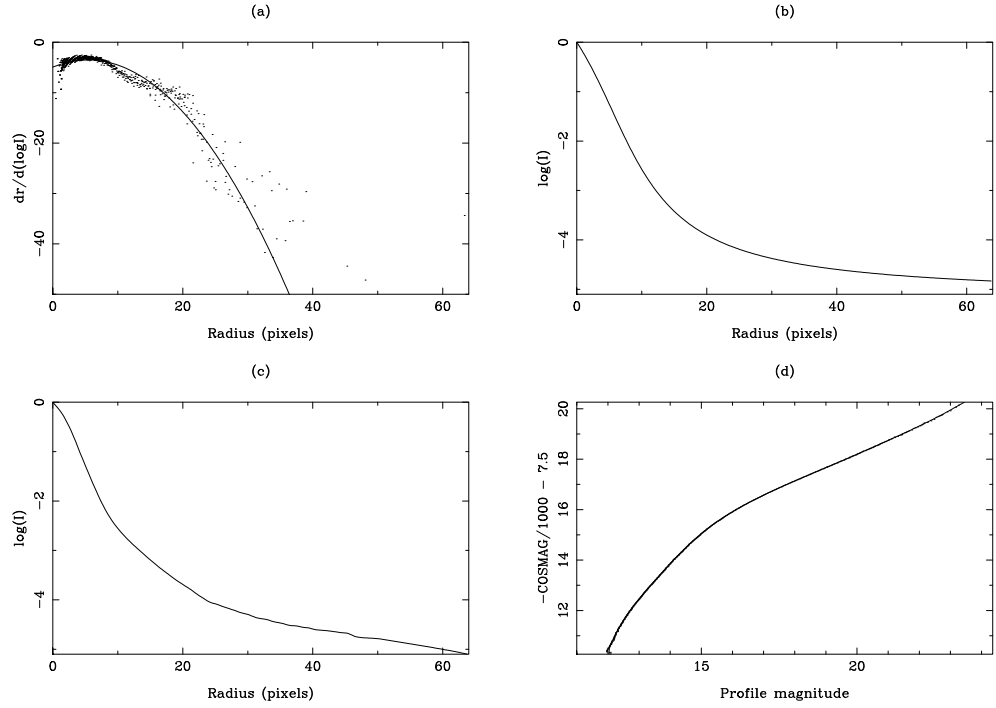}
\end{picture}
\end{center}
\caption[]{Results of the iterative determination of the stellar profile
for a typical plate: (a) differential profile estimate; (b) initial
profile; (c) iterated profile; (d) transfer function of isophotal to
profile magnitude (see Bunclark \& Irwin~1983).}
\label{apmcal}
\end{figure*}

Another part of the second stage of image classification consists of an 
iterative determination of the stellar radial profile (ie. the stellar point
spread function, PSF, for the plate) and the determination of a linear 
instrumental magnitude scale.  This follows the procedure described by 
Bunclark \& Irwin~(1983) and makes use of the fact that, to a good 
approximation, all stellar images should have the same intrinsic profile shape 
whatever their magnitude.  The areal profile measures 
(Equation~\ref{areal}) form the basis for this 
estimate, using the preset instrumental profile threshold levels as a first 
guess for the intensity scale.   Figure~\ref{apmcal} shows the results of a 
typical profile determination and linearised transfer function between 
isophotal and profile magnitude for a survey plate.  The resulting stellar 
PSF can then also be used as part of the classification scheme.

In order to make an optimal judgement about the class of an object, a 
traditional Bayesian classification scheme uses knowledge of the conditional 
and prior probabilities on the right hand side of the following equation 
\begin{equation}
p(s|d) = \frac{ p(d|s)p(s) }{p(d|s) p(s) + p(d|g) p(g) + p(d|n) p(n) } 
\end{equation}
where the probability, or likelihood, of observing the image descriptors for 
stars, galaxies and noise are given by $p(d|s)$, $p(d|g)$ and 
$p(d|n)$ respectively; $p(s)$, $p(g)$ and $p(n)$ are the 
respective prior probabilities of finding a star, galaxy or noise image; and 
$p(s|d)$ is the posterior probability of the image being a star.  In a
similar way we can write equations for $p(g|d)$ and $p(n|d)$,
the posterior probability of the image being a galaxy or noise object.
  
Unfortunately, the distribution of both galaxy and noise image parameters is 
in general unknown, implying that we cannot directly use a Bayesian estimator
in this case.  The problem is further compounded by the fact that the real
likelihood function for stellar images is also non-trivial to estimate.
This is caused by the underlying error distribution of the parameters being
non-Gaussian in the real world and also because the image parameters are 
highly correlated, with the degree of correlation varying significantly as a 
function of magnitude.

However, deriving the stellar PSF as part of the internal intensity calibration
helps to significantly reduce the number of free parameters required to 
accurately describe the shape of an object.  Each image has (up to) 8 
parameters measuring the areal profile which can be readily converted into 
the equivalent radial profile.  Clearly, images with derived radial profiles 
significantly steeper than the stellar PSF are more likely to be noise images,
while those with derived profiles much shallower than the stellar PSF are more
likey to be galaxies.  After some experimentation we found that the 
intensity-normalised slope of the sigma-weighted difference between an image 
radial profile and the stellar PSF provided a way of summarising the 
stellarness of an image using all the areal profile information. This measure 
preserves both the `sign' of the deviation and quantifies the difference.
The normalised-slope measure is then combined additively with (the relatively)
independent information derived from the intensity-weighted second moments 
and peak intensity to produce a single number measure.  Finally, as a function
of magnitude, all objects previously classified as stellar, are used to 
convert the stellarness index back to a zero-median unit-variance N(0,1) score.
This stellarness index is our pragmatic equivalent of the likelihood function 
of the data and summaries the probability that the observed data arose from
a stellar object.

\begin{figure*}
\begin{center}
\setlength{\unitlength}{1mm}
\begin{picture}(150,90)
\includegraphics{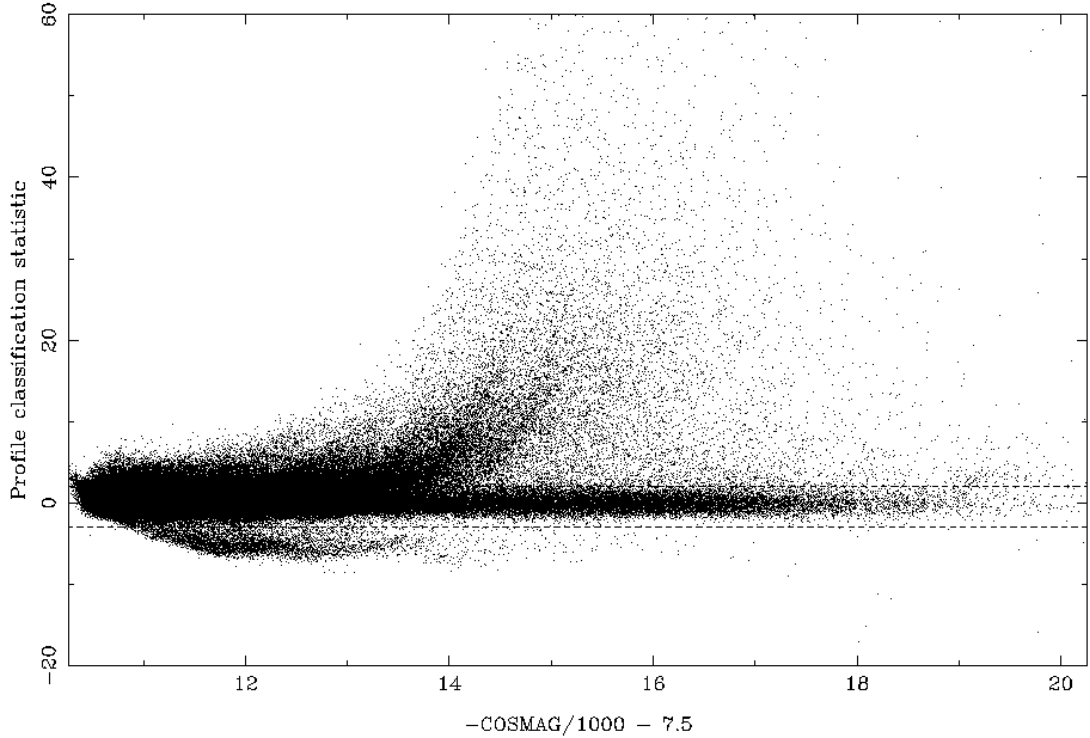}
\end{picture}
\end{center}
\caption[]{Scatter plot of the profile classification statistic
for a typical survey plate. This indicates the residuals (in terms of a
zero mean and unit standard error quantity) of each image's areal profile set 
when compared with an average stellar template. The boundaries indicate the
position of all objects finally classed as stellar. Broadly speaking,
objects with steep radial profiles (hard, noisy images) have low values,
while objects with shallow radial profiles (`fuzzy' images, more often
than not galaxies) have high values. An object matching the stellar
template perfectly has a value of zero.}
\label{reclass}
\end{figure*}

A typical example of the derived stellarness index for a survey plate is
shown in Figure~\ref{reclass} with the primary stellar locus boundaries
adopted ($+2 \sigma$ for star/galaxy and $-3 \sigma$ star/noise).
As a further refinement this average value of the stellarness index as a 
function of plate position is tabulated and used to remove small variations 
in the zero-point of the stellar locus caused by field effects.

Returning to the idealised Bayesian scheme outlined previously, for 
star/galaxy separation the main quantity of interest is 
\begin{equation}
\frac{p(s|d)}{p(g|d)} = \frac{p(s)p(d|s)}{p(g)p(d|g)}
\end{equation}
since when this ratio is greater than unity the object is classified as
a star, and when less than unity as a galaxy.  Although the underlying 
distribution, $p(d|g)$, is unknown, we do know that both galaxy and
noise images have a much wider spread of image shape descriptors than stars,
hence to a reasonable approximation $p(d|g) = U(a,b)$ and
$p(d|n) = U(c,d)$; where $U$ denotes uniform distribution and the
arguments the range over which it applies -- which in general are a function
of magnitude.
In the absence of prior information the ratio of $p(s|d)/p(g|d)$,
if $p(g|d) = U(a,b)$ and one end of the range $(a,b)$ lies within the 
normal stellar distribution, is exactly equivalent to specifying a one-sided
k-sigma boundary separating stars from galaxies. 
 
The prior information only enters via the ratio of the
expected numbers of star:to:galaxies at a particular magnitude.  For most
survey fields (i.e.~not the Galactic Plane) the total number of stellar images
and the total number of galaxy images visible on a survey plate are roughly
equal.  As a function of magnitude, however, the stellar population varies
roughly as $n(m) \propto 10^{0.25m}$ while the number of galaxy images varies 
as $n(m) \propto 10^{0.45m}$.  Using the equality of total numbers of star
and galaxies near the plate limit to define the relative normalising constant,
automatically specifies the ratio $p(s)/p(g)= constant \times 
10^{-0.2m} = 10^{-0.2(m-m_{\rm ref})}$ for all magnitudes.  
This prior information
ratio can then readily be incorporated into the previous boundary definition 
by defining a new k'-sigma boundary using the assumed N(0,1) distribution of 
the likelihood function to convert ratios to sigmas.

Although inevitably somewhat adhoc, this scheme does allow us to fold in 
additional information within the framework of a Bayesian approach 
(e.g.~prior knowledge of star:to:galaxy ratios and the knowledge that galaxies
and noise images have a wide range of image shapes).  The advantages of
decision making directly in k-sigma space are that it is straightforward to 
place realistic constraints on the influence of the prior
knowledge, i.e.~by limiting
how much the boundary can shift, and it also mitigates against the effect of
non-Gaussian errors by automatically scaling the boundary using a robust
sigma estimator.


\subsection{Photometric calibration}

The isophotal instrumental magnitude scale for IAM data is non--linear with
respect to external apparent magnitude. Non--linearities are introduced by,
for example, thresholding and the limited dynamic range of the measuring
machine in addition to any non--linearities inherent in the photographic
characteristic curve with respect to the linear $D-\log I$
scale assumed. Additionally,
the magnitude scale for galaxies will be, in general, different to that of
stars. The reason for this is that galaxies are generally more extended and
of lower surface brightness than stars, and as such have integrated magnitudes
that are less susceptible to any dynamic range limit during measurement.
Moreover, the linearised profile magnitude for stars is easier to calibrate than
the isophotal scale when external calibration data are limited since it is
better behaved. Note that the profile magnitude for a non--stellar object is not
useful since such objects have, of course, non--stellar profiles. For these
reasons, the stellar/non--stellar magnitude calibrations for the survey data
are treated separately. The calibration procedure adopted was as follows:

\subsubsection{Gradient of the instrumental stellar profile magnitude scale}
\label{grads}

\begin{figure*}
\begin{center}
\setlength{\unitlength}{1mm}
\begin{picture}(150,90)
\includegraphics{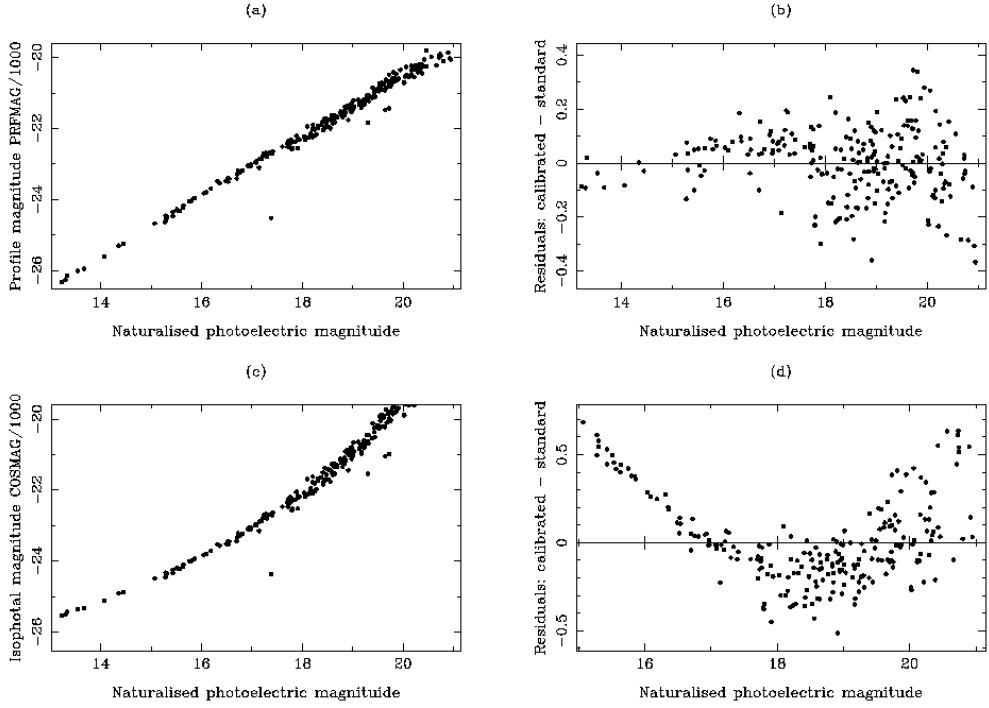}
\end{picture}
\end{center}
\caption[]{Plots of naturalised photoelectric magnitudes verus (a) profile
magnitude and (c) isophotal magnitude for R~band data in survey field~413.
Panels (b) and~(d) show the non--linear
components after linear least--squares regression fits. The profile scale
is closely linear over $\sim8$~mags. The isophotal scale is non--linear
at the level of $>0.5$~mag; furthermore, such non--linearities are not
repeatable from field to field.}
\label{gradmeas}
\end{figure*}

For each of the survey colours (e.g.~SERC--J, ESO--R) the gradient of the
faintest 7~mag of the profile magnitude scale was measured against CCD data
from Boyle, Shanks \& Croom~(1995) and Croom et al.~(1999). Throughout this
paper, where any comparisons have been done between photographic and
photoelectric photometry, colour equations from Blair \& Gilmore~(1982),
Bessell~(1986) and Evans~(1989) have been used to naturalise the magnitudes
to the photographic 
B$_{\rm J}$/R$_{\rm 63F}$/R$_{\rm 59F}$/E$_{2224}$/I$_{\rm IVN}$ systems.
The results of
the measurements are shown in Table~\ref{gradients}, where the number of
plate gradients and a standard error estimate are also given.
\begin{table}
\begin{center}
\begin{tabular}{crrrc}
Survey & No.~of & \multicolumn{3}{c}{Gradient of profile scale} \\
       & fields  & min & max & mean~$\pm\sigma$ \\
\multicolumn{5}{c}{ }\\
SERC--J/EJ      & 25 & 1.133 & 1.456 & $1.281\pm0.080$ \\
AAO--R/SERC--ER & 19 & 1.118 & 1.239 & $1.178\pm0.028$ \\
SERC--I         &  6 & 0.947 & 1.156 & $1.086\pm0.074$ \\
ESO--R          & 14 & 1.038 & 1.179 & $1.107\pm0.044$ \\
POSS--I E       & 11 & 0.956 & 1.212 & $1.070\pm0.081$ \\
AAO--R on 4415 film &  4 & 1.182 & 1.297 & $1.243\pm0.044$ \\
\end{tabular}
\caption[ ]{Gradients of the profile magnitude scale for the various types of
survey plate comprising the SGC dataset.}
\label{gradients}
\end{center}
\end{table}
We found that the RMS residuals of the linear fits between the profile and
standard magnitudes were in the range~$0.1^{\rm m}$ to~$0.2^{\rm m}$, in
line with expectations for photographic photometry of fainter stars and also
demonstrating that the profile magnitude scale has no increased scatter over
that present in the raw isophotal scale (e.g.~Bunclark \& Irwin~1983). The 
mean gradients in Table~\ref{gradients} are reasonably close to unity with
small standard errors. This implies that the assumed calibration used for
transmission to intensity was a reasonably good one. Furthermore, the small
scatter in the gradients means that rough calibration to the plate limit is
possible on a field--by--field basis using a small number of relatively bright
standards in each field to define a bright stellar calibration curve that can
be extrapolated to the plate limits. Figure~\ref{gradmeas} 
shows a typical example
of a deep photometric sequence plotted against the linearised profile magnitude
scale, demonstrating how closely linear the profile scale is over
many magnitudes above the plate limit. Contrast this with the same plotted for
the isophotal scale, which is non--linear at the 0.5~mag level.

\subsubsection{Rough profile magnitude calibration for stars}
\label{rough}

\begin{figure*}
\begin{center}
\setlength{\unitlength}{1mm}
\begin{picture}(150,90)
\includegraphics{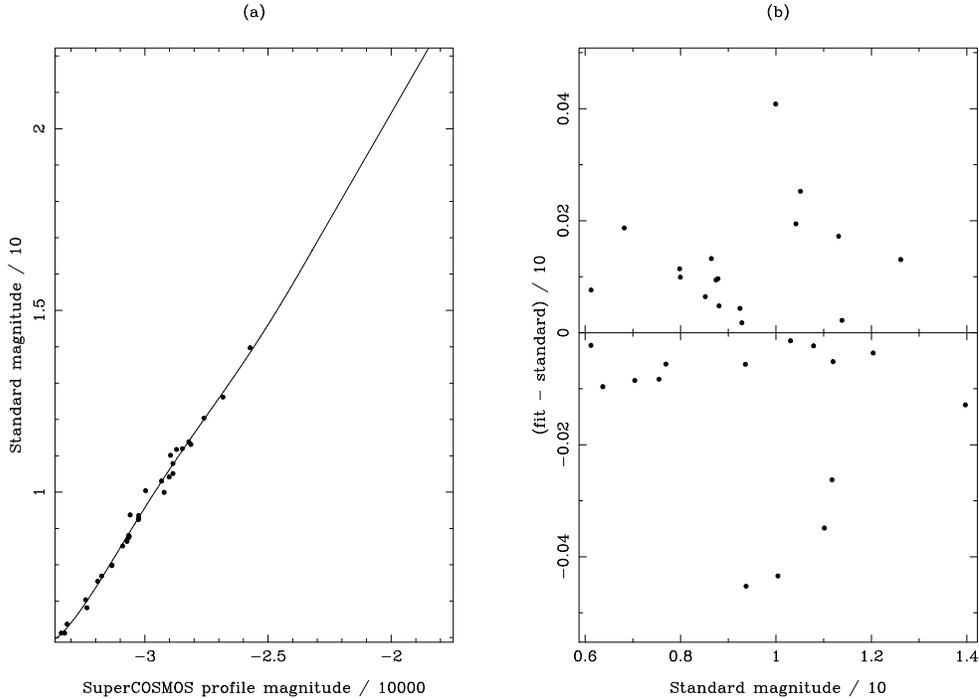}
\end{picture}
\end{center}
\caption[]{Calibration curve of the linearised stellar profile magnitude
scale versus external photoelectric photometry naturalised to the
photographic system. The solid line in (a) is a least--squares polynomial fit
constrained to have the required gradient at the faint end. In (b) we show the
residuals of the data about the fit.}
\label{ccurve}
\end{figure*}

In order to roughly calibrate the stellar profile magnitude scale of
each field with respect to external standards, a compilation of
BV photometry over the whole sky was made by combining the Tycho Catalogue
(ESA~1997) and the Guide Star Photometric Catalogue~{\sc I} (Lasker et
al.~1988). This provided two colour photometry down to V~$\sim15$ in every 
survey field. Standard magnitudes in the RI bands have been estimated from
relationships between (B--V) versus (V--R) and (V--I) derived from the
multicolour standard star data of Landolt~(1992). For each field, a calibration
curve of the profile instrumental magnitude versus naturalised standard
magnitude was derived. Figure~\ref{ccurve} shows a typical example for the
survey data. The solid lines are least--squares fits to the data, with the
added linear constraints that at a fixed point 3~mag fainter than the the
faintest calibrating standard, the gradient is set at the appropriate mean
value given in Table~\ref{gradients} while the second derivative of the
fitting function is set to zero. In this way, the calibrating function can be
linearly extrapolated from this point to the plate limit, and the calibrating
function has no discontinuities in itself or its first derivative. This 
somewhat complicated procedure is an attempt to realistically model the shape
of the `linearised' stellar profile magnitude scale, which is indeed linear
over its faintest 7~mag range (e.g.\ Figure~\ref{gradmeas}), 
but which departs from linearity brighter than
this due to the fundamental dynamic range limits of the machine as dictated by
diffraction--limited performance in the imaging optics and
the transmission digitisation range, along with the dynamic range limit
of the photographs themselves.

The fact that the calibrations are extrapolated for the fainter magnitude
range coupled with the paucity of external calibrating stars on the linear
part of the profile scale results in small (10\% to~20\%) errors in the
zeropoint of the faint magnitude calibration of each field. This problem is
dealt with in the next Section. 

\subsubsection{Global zeropoint definition}
\label{global}

All the Schmidt surveys have varying degrees of overlap between adjacent fields.
For the POSS--I E and ESO--R surveys, this overlap can be very small (or 
non-existent in the case of a few adjacent ESO--R fields), but for the 
UK~Schmidt surveys the field centres were chosen to give $\sim0.5^{\circ}$
overlap strips for all fields (this system of field centres, reflected into
the northern hemisphere, has been adopted for the POSS--II survey). These
overlap regions can be used to define a set of individual field photometric
zeropoints such that the mean difference over all stars between the stellar
magnitudes from the two measurements available per star in any given overlap
region is zero. Consider any two adjacent fields~$m$ and~$n$. The mean
(systematic) difference between the stellar magnitudes in the overlap region
is $\Delta_{nm}=z_n-z_m$ where $z_n$ and $z_m$ are the unknown zeropoints.
A mosaic of many fields gives a (singular) set of equations in the $\Delta$
and $z$ values. Such a set may be solved by singular value decomposition to
yield minimal zeropoints that solve the set. However, we have used faint
CCD photometry to externally fix as many zeropoints as possible over the
survey region. This makes the equation set non--singular and yields a unique
vector of zeropoints that are tied to an external system. This procedure is
similar to that adopted by Maddox, Efstathiou \& Sutherland~(1990) for the APM
Galaxy Survey except that we have {\em not} solved for the field--to--field
differences in the gradients of the photometric scales. Also, our sources
of external CCD photometry included data from the Guide Star Photometric
Catalogue~II (Postman et al.~1997), Boyle et al.~(1995), Croom et
al.~(1999), Cunow et al.~(1997) and references therein, unpublished data used
in the calibration of the Edinburgh--Durham Southern Galaxy Catalogue 
(e.g.~Nichol~1993 and references therein) and Stobie, Sagar \& Gilmore~(1985)
in addition to that published in Maddox et al.~(1990). At the time of
writing, approximately 50\% of all southern hemisphere fields have external
CCD data defining their photometric zeropoint.

\subsubsection{Correction of systematic errors in stellar colours as a function
of magnitude and plate position}
\label{colbias}

\begin{figure}
\begin{center}
\setlength{\unitlength}{1mm}
\begin{picture}(80,130)
\includegraphics{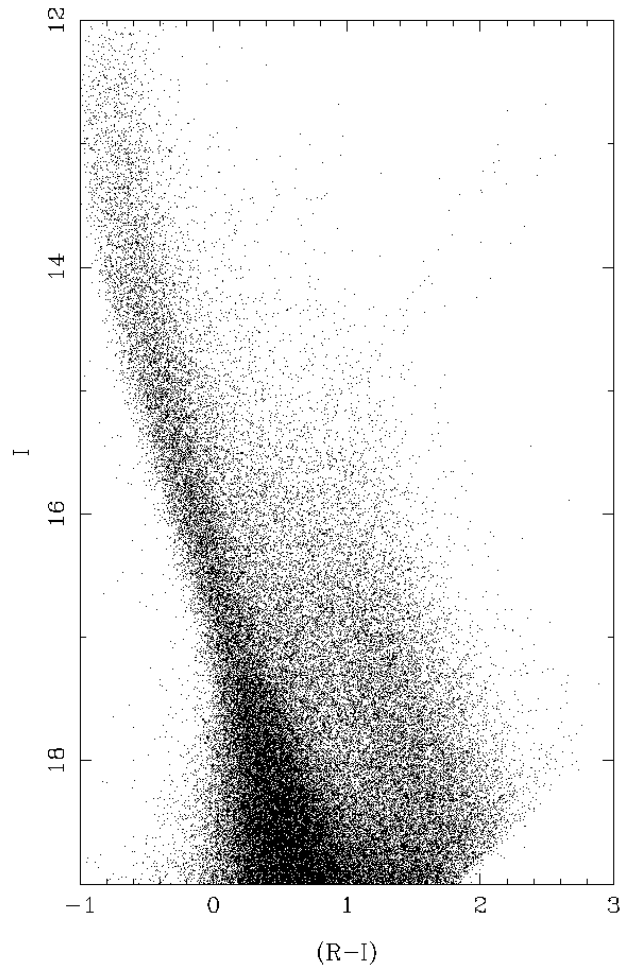}
\end{picture}
\end{center}
\caption[]{Stellar colour--magnitude diagram (I versus R--I) for survey 
field~758 {\em before} correction of systematic colour errors as a function 
of magnitude and position. There is no strong reddening in this high 
latitude field, so the large colour change with magnitude has to be due
to systematic calibration errors.}
\label{cmdbefore}
\end{figure}

\begin{figure}
\begin{center}
\setlength{\unitlength}{1mm}
\begin{picture}(80,80)
\includegraphics{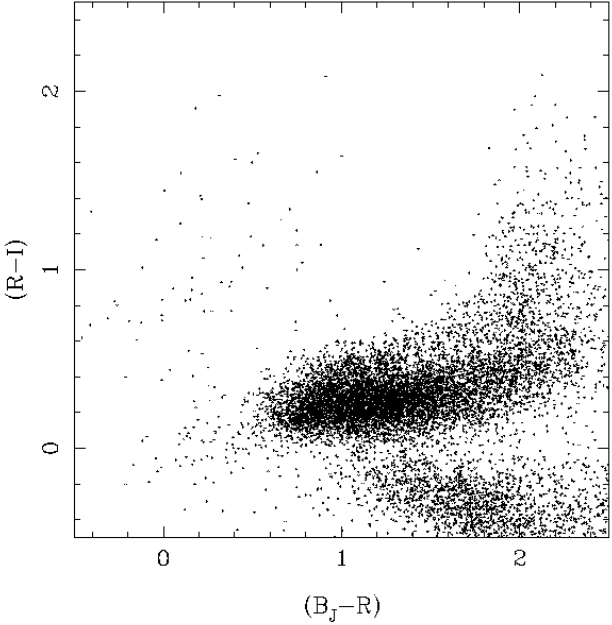}
\end{picture}
\end{center}
\caption[]{Stellar two--colour diagram (B$_{\rm J}$--R versus R--I in the
magnitude range $17<{\rm B}_{\rm J}<18$) 
for 9 contigous survey fields around the South Galactic Pole
{\em before} correction of systematic colour errors as a function of magnitude
and position.}
\label{twocolbefore}
\end{figure}

The photometric calibration of photographic plates based on calibrating stars
necessarily assumes that there exists a unique calibration curve that is
applicable over the entire plate. In general, this is {\em not} the case.
For example, graphs of external photoelectric magnitude versus internal
linearised profile magnitude, as a function of plate position, show significant
changes in gradient. One possible cause of this is that the response of the
photographic emulsion is not uniform over the plate -- i.e.~the 
characteristic curve of the emulsion is not constant. Whatever the cause, the
results are the same: systematic errors in photometry as a function of magnitude
and position occur, and those errors manifest themselves most visibly when 
magnitudes are combined to generate colour indices. Moreover, we have
assumed a constant value of the gradient of the profile magnitude scale in
any given passband, and as Table~\ref{gradients} shows there is a significant
dispersion about the mean values adopted which adds to any systematic effects
within a given plate. For example, Figure~\ref{cmdbefore} shows an~I versus
(R--I) colour--magnitude diagram for survey field~758. There is apparently
a trend for brighter stars to get become bluer in this diagram, along with
a significant increase in scatter about the modal colour for brighter 
magnitude ranges.
These effects are both due to systematic errors in photometry as a function
of magnitude and position. Furthermore, Figure~\ref{twocolbefore} shows a
stellar (J--R) versus~(R--I) two--colour diagram for a mosaic of
9~fields at the South Galactic Pole 
for the magnitude range $17<{\rm B}_{\rm J}<18$. While the
general trend of red objects being red in both colours is apparent, it is
also true that stars having intermediate colours do not show a good
correlation between (J--R) and~(R--I). The scatter in Figure~\ref{twocolbefore}
is rather large, and masks astrophysically useful information.

\begin{figure}
\begin{center}
\setlength{\unitlength}{1mm}
\begin{picture}(80,130)
\includegraphics{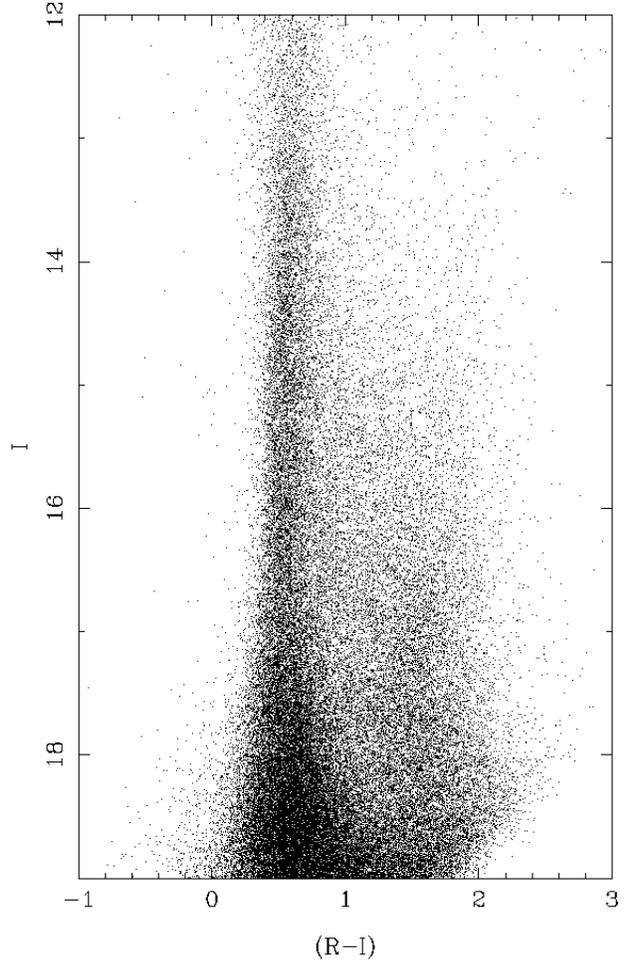}
\end{picture}
\end{center}
\caption[]{Stellar colour--magnitude diagram (I versus R--I) 
for survey field~758
{\em after} correction of systematic colour errors as a function of magnitude
and position.}
\label{cmdafter}
\end{figure}

\begin{figure}
\begin{center}
\setlength{\unitlength}{1mm}
\begin{picture}(80,80)
\includegraphics{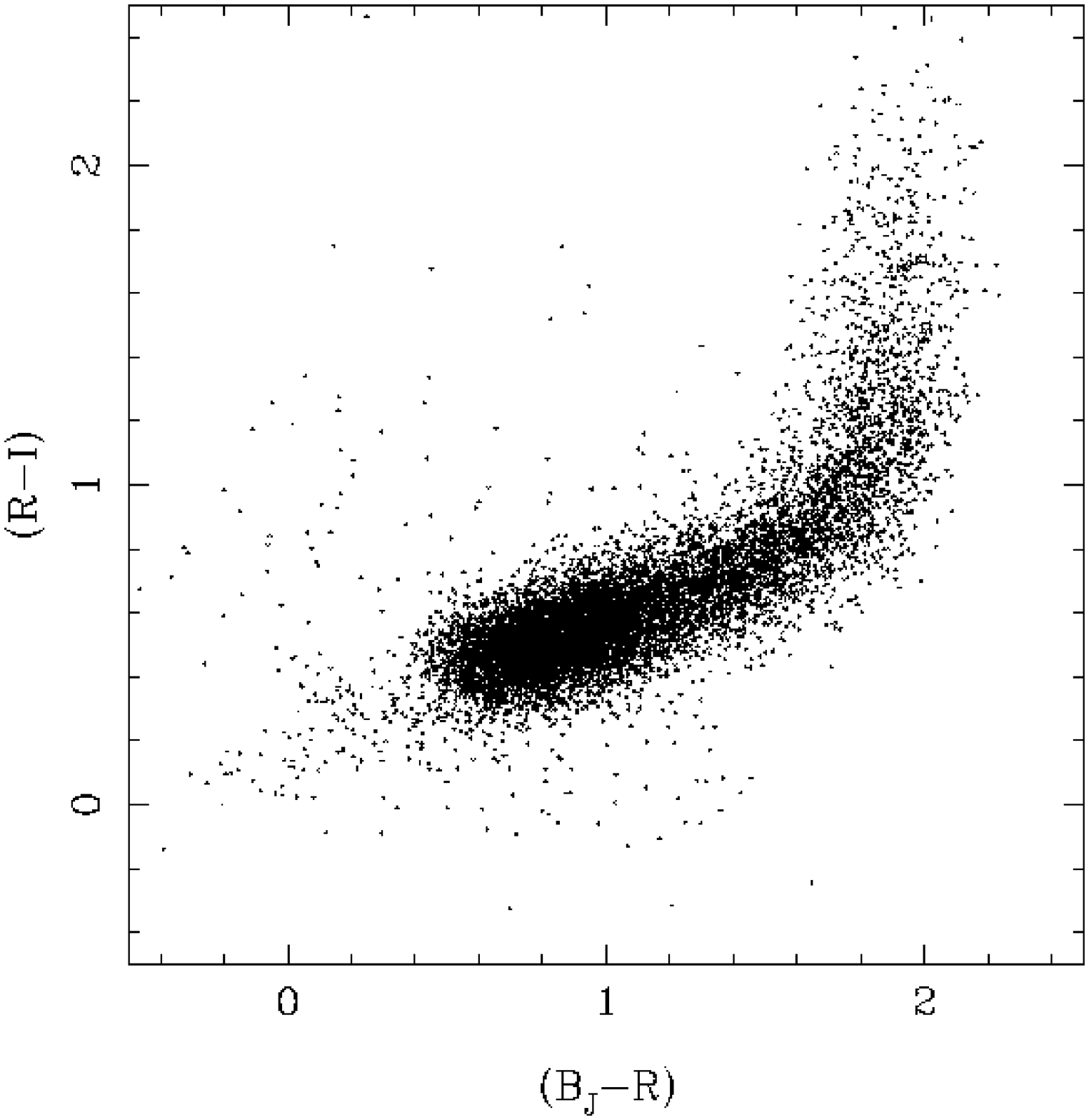}
\end{picture}
\end{center}
\caption[]{Stellar two--colour diagram (B$_{\rm J}$--R versus R--I in the
magnitude range $17<{\rm B}_{\rm J}<18$)
for 9 contigous survey fields around the South Galactic Pole
{\em after} correction of systematic colour errors as a function of magnitude
and position.}
\label{twocolafter}
\end{figure}

These problems have been documented in the past (e.g.~Hambly et al.~1995;
Goldschmidt \& Miller~1998 and references therein). In the abscence of
large--scale external calibration sequences, there is no option other than
to use the stellar distributions themselves (in colour, as a function of 
magnitude and position) to measure the size of these effects and to calibrate
them out. Such corrections require that one passband is chosen as the 
standard with respect to which all other passbands are corrected. Here, we 
have chosen the J~plates for this purpose because the most widely available
external photometry is in the B~band and we expect that the rough calibrations
(Section~\ref{rough}) will be the most accurate in this passband. The
correction procedure is straightforward, and is applied to each set of
R~and~I field data individually. First, shifts in colour as a function of
position are measured by cross--correlating smoothed histograms of
number--colour counts as a function of magnitude and plate position in coarse
bins of 1~mag and 2~cm respectively. The resulting data cubes of shifts are
then padded, resampled onto a $3\times$ finer grid and linearly filtered and
smoothed to yield final shift values to be applied to the data. The final
stage of the correction procedure consists of forming number--magnitude
histograms as a function of magnitude only, in 0.1~mag bins, and finding the
10th percentile blue `edge' of the distributions. Hence, a shift of this value
with respect to a global value (measured as the median 10th percentil blue
edge of faint stars over all the fields in the survey) is measured and can be
applied to the data to shift and straighten the blue edge to an internally
consistent position. Table~\ref{blueshifts} shows the values measured over
the whole survey for anchoring the colours in this way. 
Figures~\ref{cmdafter} and~\ref{twocolafter} show the results of this
procedure for direct comparison with Figures~\ref{cmdbefore} 
and~\ref{twocolbefore}.  There is a significant improvement in the
morphology and reduction in scatter in these diagrams after correction.
\begin{table}
\begin{center}
\begin{tabular}{ccc}
Survey & Colour & Position of\\
       &        & blue edge\\
\multicolumn{3}{c}{ }\\
AAO--R/SERC--ER & (B$_{\rm J}$--R) & 0.789 \\
SERC--I         & (B$_{\rm J}$--I) & 1.204 \\
ESO--R          & (B$_{\rm J}$--R) & 0.774 \\
\end{tabular}
\caption[ ]{Values of the 10th--percentile blue `edge' used in fixing the
survey colours (see text).}
\label{blueshifts}
\end{center}
\end{table}

Of course, it should be noted that this procedure can potentially remove real
astronomical features from colour--magnitude and two--colour diagrams. For
example, differential reddening across a field would tend to be calibrated
out by the above procedure. For this reason, the database access software
(Paper~{\sc I}) allows the user the option of turning off these corrections
at access time so that uncorrected colours are available if required.

\subsubsection{Galaxy (isophotal) magnitude calibration}

As mentioned previously, the galaxy magnitude calibration must be treated
separately to that of stellar objects. The isophotal magnitude scale is
tied to that of stellar objects using the faintest $2.5^{\rm m}$ range of
objects where the isophotal scale for stars and galaxies is the same since
at these magnitudes, the majority of
galaxies are unresolved on the plates and have the same
profile shapes. In order to apply a simple linear fit between isophotal and
profile magnitudes, however, the isophotal scale must first be corrected for
thresholding non--linearity (the profile scale is, of course, linear over
much more than the faintest $2.5^{\rm m}$ range). 

\begin{figure*}
\begin{center}
\setlength{\unitlength}{1mm}
\begin{picture}(150,90)
\includegraphics{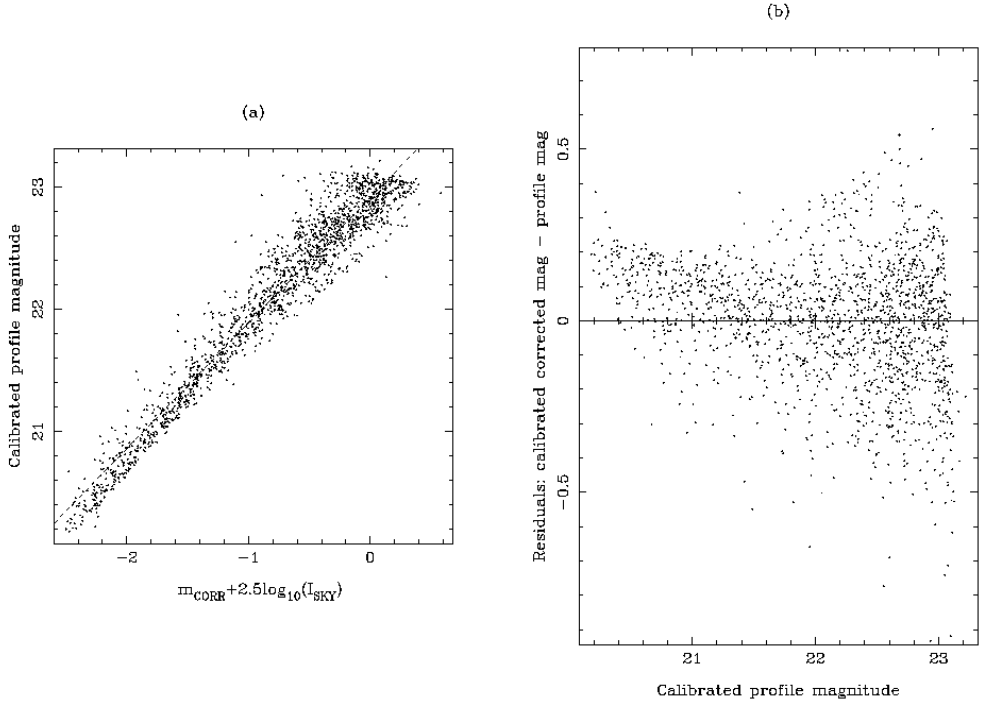}
\end{picture}
\end{center}
\caption[]{Calibration curve of the m$_{\rm CORR}$ (see text) for non--stellar
objects versus calibrated profile magnitude for faint, unresolved images.}
\label{galcal}
\end{figure*}

The thresholding correction to the isophotal scale was modelled assuming a
Gaussian profile,
\begin{equation}
I(r)=I_{\rm PEAK}\exp(-r^2/2\sigma^2), 
\end{equation}
from which integration yields $I_{\rm TOT}=2\pi\sigma^2I_{\rm PEAK}$ where
$\sigma$=FWHM of the profile. If thresholding truncates the integrated 
brightness at a radius $r_t$, then it is easy to show that
\begin{equation}
r_t^2=2\sigma^2\log_{\rm e}\left\{\frac{100I_{\rm PEAK}}{I_{\rm SKY}\times{\rm PCUT}}\right\}, 
\end{equation}
where $I_{\rm SKY}$ and PCUT are defined by Equation~\ref{threshold} in
Section~\ref{thresholding}. We also have that the difference between the total
and isophotal integrated intensities of an object is given by
\begin{eqnarray}
I_{\rm TOT}-I_{\rm ISOPHOTAL} & = & 2\pi 
I_{\rm PEAK}\int_{r_t}^{\infty}\exp(-r^2/2\sigma^2)r{\rm d}r \nonumber \\
                              & = & 
\frac{2\pi\sigma^2I_{\rm SKY}\times{\rm PCUT}}{100} 
\end{eqnarray}
and defining an isophotal correction $\epsilon$ such that
\begin{equation}
{\rm m}_{\rm CORR}={\rm m}_{\rm ISOPHOTAL}-2.5\log_{10}(1+\epsilon), 
\end{equation}
it is straightforward to show that
\begin{equation}
\epsilon=
\left\{1-\frac{I_{\rm SKY}\times{\rm PCUT}}{100I_{\rm PEAK}}\right\}^{-1} -1.
\end{equation}
Hence, a calibration of m$_{\rm CORR}$ versus the calibrated profile magnitude
for the same image is possible, and is furthermore applicable to the full
threshold corrected isophotal scale. Note that $\epsilon$ is not a correction
to `total' galaxy magnitudes (e.g.~Young et al.~1998 and references therein)
since such a correction is dependent on galaxy profile shape. Note also that
we calibrate m$_{\rm CORR}+2.5\log_{10}I_{\rm SKY}$ to provide a first--order
correction for vignetting and differential desensitisation as a function of
plate position (e.g.~Maddox et al.~1990). An example calibration of 
m$_{\rm CORR}$ versus calibrated profile magnitude is shown in 
Figure~\ref{galcal}. In this way a calibration for non--stellar objects is
defined in every field, this calibration being tied to the same external
zeropoints as the stellar magnitude scale.

It is worth emphasising that the calibrations set up for the survey photometry
take the form of coefficients values that are applied to raw measurements as
and when the data are accessed (see Paper~{\sc I}). This allows recalibration
at any point in the future when better methods and/or better external
calibration data become available. The raw, uncalibrated and uncorrected 
isophotal and profile instrumental magnitudes are also available, along with
all morphological information produced during pixel analysis. This will allow,
for example, investigation and implementation of more sophisticated galaxy
calibration techniques in the future (e.g.~Young et al.~1998 and references
therein).

\section{Results and discussion}
\label{results}

When assessing the accuracy of the measurement, classification and calibration
schemes detailed in the previous Section, it is important to distinguish
between the various types of error that it is possible to measure and quote.
This is particularly important when comparing the SSS characteristics against
those of other survey programmes (e.g.~Paper~{\sc I}). Here, we have assessed
the internal consistency of the survey data by comparison between data for the
same objects from overlap regions (i.e.~where two independent measurements of
the same quantity are available). Under the assumption that the two overlapping
plates have the same characteristics of image quality and signal--to--noise
ratio, this also provides estimates of the random errors (dividing by
$\surd 2$, where appropriate, for a single measurement). 
Note that similar comparisons in overlap regions other than those specifically
used here yield similar results provided they have Galactic latitudes
$|b|\geq30^{\circ}$.
As it will be shown in the next Sections, the random errors in single colour
photometry are dominated by position
and magnitude dependent systematic errors for brighter magnitude ranges, and it 
is therefore useful to quote an apparent `standard' error in the presence of
such effects, and also a relative error when comparing photometry within
restricted magnitude and position ranges. The latter thus reflects the true,
underlying random errors in the measurement process and is useful when the data
are being used in isolation (for example, inter--comparison of object 
magnitudes)
while the former is more appropriate when measurements are taken at face--value
and compared with external data. 
Note however that the calibration procedure removes systematic
errors in colour as a function of magnitude and position {\em and} across
the survey as a whole (Section~\ref{colbias}) so that the SSS object colours
(e.g.~B$_{\rm J}$--R, R--I) are much more accurate than the external errors
on individual passband magnitudes. This is illustrated later.

Of course, systematic errors can only be truly
assessed with respect to external data that can be regarded as `gold' standard,
i.e.~free from systematic error and having negligible random error, and we also
describe comparisons against external CCD data that are of sufficient angular
resolution and signal--to--noise that they can be regarded as error--free for
these purposes. 

\subsection{Internal consistency and relative accuracy}

\subsubsection{Image classification}

\begin{figure*}
\begin{center}
\setlength{\unitlength}{1mm}
\begin{picture}(150,90)
\includegraphics{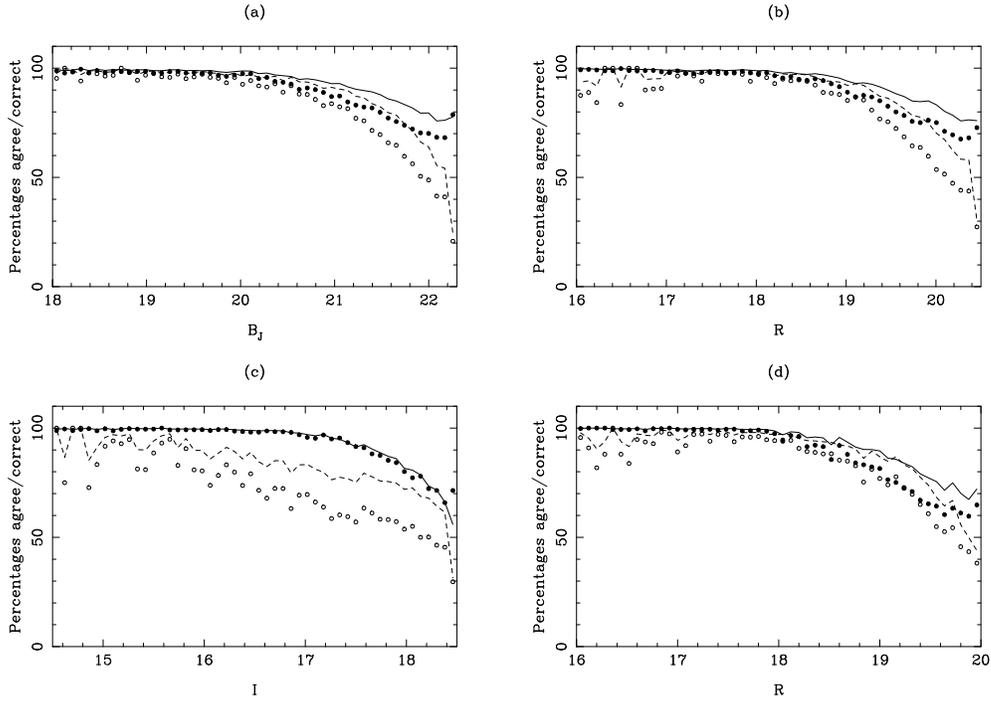}
\end{picture}
\end{center}
\caption[]{Classification scores when comparing data from two overlapping
plates for (a) SERC--J/EJ, (b) AAO--R/SERC--R, (c) SERC--I and (d) ESO--R
surveys. Solid circles are for stars while open circles are for galaxies;
solid and dashed lines indicate the percentage accuracy of a single
classification for stars and galaxies respectively (see text).}
\label{clsprob}
\end{figure*}

In order to assess the consistency of the image classification, we compared data
from each of the four survey types J,R,I and ESO--R in the large 
($\sim4^{\circ}$) overlap region between field numbers~292 and~241. In 
Figure~\ref{clsprob}(a) to~(d) we show plots of the percentage agreement between
the two independent classifications, as a funcion of magnitude, for stars
(solid circles) and galaxies (open circles).
These were analysed for the {\em underlying} correct classification for a
single measurement using the following argument. Assume that there are just
the two image classes, stellar and galaxian (the fraction of noise and
unclassifiable images on any plate is always $\ll 1$\%). If we denote the
probability that a real star is classified as such in any given magnitude
range by $p_1(s|s)$ and the probability that a real galaxy is classed as such
by $p_1(g|g)$ along with probabilities for incorrect classification by
$p_1(s|g)$ and $p_1(g|s)$ (where the subscript refers to plate~1 and similar
definitions are made for the overlapping
plate~2), we have that for plate~1 for example
\begin{eqnarray}
\label{one}
p_1(s|s)+p_1(g|s) & = & 1 \\
\label{two}
p_1(g|g)+p_1(s|g) & = & 1 
\end{eqnarray}
and similarly for plate~2. Between the two independent measurements, the 
probability of agreement for stellar class (for example) is equal to the sum of
the probability that both measurements class a real star as such and the 
probability that both measurements misclassify a galaxy as a star:
\begin{equation}
\label{three}
p({\rm agree}\ s)=p_1(s|s)p_2(s|s)+p_1(s|g)p_2(s|g);
\end{equation}
likewise for agreement on galaxian classification:
\begin{equation}
\label{four}
p({\rm agree}\ g)=p_1(g|g)p_2(g|g)+p_1(g|s)p_2(g|s).
\end{equation}
Assume now that the noise parameters and general image quality
of the plates are the same, i.e.~write
$p(s|s)=p_1(s|s)=p_2(s|s)$ etc.~and since we are not interested in the cross
terms in Equations~\ref{three} and~\ref{four} they can be eliminated using
Equations~\ref{one} and~\ref{two} to yield
\begin{eqnarray}
p({\rm agree}\ s) & = & p(s|s)^2+\{1-p(g|g)\}^2, \\
p({\rm agree}\ g) & = & p(g|g)^2+\{1-p(s|s)\}^2, 
\end{eqnarray}
which is a set of two non--linear equations in the two unknowns~$p(s|s)$
and~$p(g|g)$. These are soluble using a simple iterative scheme, and in
Figure~\ref{clsprob}(a) to~(d) we plot the derived functions $p(s|s)$
(solid lines) and $p(g|g)$ (dashed lines).

From Figure~\ref{clsprob}, the image classification scheme is clearly working
well, with stellar classification being {\em internally} repeatable/reliable
at the $>95$\% level for
magnitudes B$_{\rm J}<21$, R~$<19$ (or R~$<18.5$ for ESO--R) and I~$<17.5$.
For galaxies, classification is reliable at the $>90$\% level in the ranges
$17.5<{\rm B}_{\rm J}<21.5$, $15<{\rm R}<19$ but is clearly not so good for the
I plates. This is possibly due to the generally poorer signal--to--noise of I~band
Schmidt plates. Note that reliable image classification for all I~images is of
course available from the B$_{\rm J}$ data unless an object is severely
reddened or intrinsically very red. It is important to 
emphasise that the only true
test of the image classification reliability is against external data. This is
because the above analysis cannot take into account any systematic effects
(for example, classification of large numbers of faint, unresolved galaxies
as stellar objects). Such a rigorous test is detailed in Section~\ref{extsg}.
On the other hand, this internal test illustrates that the classification
scheme is not severely limited by random errors at magnitudes as faint as
$\sim1.5^{\rm m}$ above the respective plate limits.

\subsubsection{Photometry (single passband)}

Using the same overlap region as in the previous Section, we have compared the 
photometry to assess the internal accuracy of the photometric measurements and 
calibrations. Tables~\ref{int_j} to~\ref{int_s} give detailed accuracy figures
as a function of magnitude for the J,R,I and ESO--R surveys respectively. In
each Table, the figures for stars and galaxies are given separately since
galaxy photometry will be less accurate at a given magnitude (e.g.~Irwin 1985).
The columns in each Table for each image type are the number of objects used
in obtaining the measurements; the raw RMS (including any systematic effects);
the median global zeropoint error; the RMS when this global zeropoint is 
removed; and finally the RMS when a position--dependent zeropoint is removed.
Zeropoints as a function of position were determined from the data by taking
median values over a scale size of $\sim2$~cm ($\sim20$~arcmin) on the plates.
\begin{table*}
\begin{center}
\begin{tabular}{crccccrcccc}
 & \multicolumn{5}{c}{Stars} & \multicolumn{5}{c}{Galaxies} \\
Range & \multicolumn{1}{c}{N} & $\sigma_{\rm RAW}$ & ZP & $\sigma_{\rm -ZP}$ & $\sigma_{\rm REL}$  
& \multicolumn{1}{c}{N} & $\sigma_{\rm RAW}$ & ZP & $\sigma_{\rm -ZP}$ & $\sigma_{\rm REL}$ \\
\multicolumn{11}{c}{ }\\
$  9.0 <{\rm B}_{\rm J}< 10.0$&   51 &  0.55  & $-0.53$ &   0.13 & --- & --- & --- & --- & --- & --- \\
$ 10.0 <{\rm B}_{\rm J}< 11.0$&   92 &  0.52  & $-0.49$ &   0.10 & --- & --- & --- & --- & --- & --- \\
$ 11.0 <{\rm B}_{\rm J}< 12.0$&  197 &  0.40  & $-0.38$ &   0.09 & --- & --- & --- & --- & --- & --- \\
$ 12.0 <{\rm B}_{\rm J}< 13.0$&  438 &  0.36  & $-0.35$ &   0.11 & 0.03 & --- & --- & --- & --- & --- \\
$ 13.0 <{\rm B}_{\rm J}< 14.0$&  789 &  0.37  & $-0.36$ &   0.13 & 0.03 & --- & --- & --- & --- & --- \\
$ 14.0 <{\rm B}_{\rm J}< 15.0$& 1384 &  0.43  & $-0.41$ &   0.13 & 0.03 & --- & --- & --- & --- & --- \\
$ 15.0 <{\rm B}_{\rm J}< 16.0$& 2014 &  0.47  & $-0.45$ &   0.14 & 0.04 &  16  & 0.40 & 0.38 & 0.12 & ---\\
$ 16.0 <{\rm B}_{\rm J}< 17.0$& 3105 &  0.46  & $-0.44$ &   0.14 & 0.04 & 115  & 0.37 & 0.35 & 0.06 & ---\\
$ 17.0 <{\rm B}_{\rm J}< 18.0$& 4305 &  0.42  & $-0.40$ &   0.14 & 0.05 & 505  & 0.27 & 0.24 & 0.08 & 0.08\\
$ 18.0 <{\rm B}_{\rm J}< 19.0$& 5595 &  0.34  & $-0.32$ &   0.12 & 0.05 &1522  & 0.19 & 0.16 & 0.07 & 0.06\\
$ 19.0 <{\rm B}_{\rm J}< 20.0$& 7831 &  0.25  & $-0.23$ &   0.12 & 0.06 &4106  & 0.12 & 0.09 & 0.07 & 0.06\\
$ 20.0 <{\rm B}_{\rm J}< 21.0$&10851 &  0.17  & $-0.14$ &   0.12 & 0.07 &9509  & 0.10 & 0.01 & 0.10 & 0.09\\
$ 21.0 <{\rm B}_{\rm J}< 22.0$&18679 &  0.16  & $-0.07$ &   0.15 & 0.12 &13563 & 0.17 & 0.05 & 0.17 & 0.16\\
$ 22.0 <{\rm B}_{\rm J}< 23.0$&27806 &  0.20  & $-0.03$ &   0.20 & 0.18 &6429  & 0.27 & 0.08 & 0.27 & 0.25\\
\end{tabular}
\caption[ ]{Internal systematic and random errors measured from the overlapping plate
pair SERC--J fields~292 and~241.}
\label{int_j}
\end{center}
\end{table*}
\begin{table*}
\begin{center}
\begin{tabular}{crccccrcccc}
 & \multicolumn{5}{c}{Stars} & \multicolumn{5}{c}{Galaxies} \\
Range & \multicolumn{1}{c}{N} & $\sigma_{\rm RAW}$ & ZP & $\sigma_{\rm -ZP}$ & $\sigma_{\rm REL}$  
& \multicolumn{1}{c}{N} & $\sigma_{\rm RAW}$ & ZP & $\sigma_{\rm -ZP}$ & $\sigma_{\rm REL}$ \\
\multicolumn{11}{c}{ }\\
$   9.0 < {\rm R} < 10.0$ &    89 & 0.56 & 0.54 & 0.06 & --- & --- & --- & --- & --- & --- \\
$  10.0 < {\rm R} < 11.0$ &   140 & 0.45 & 0.43 & 0.05 & --- & --- & --- & --- & --- & --- \\
$  11.0 < {\rm R} < 12.0$ &   346 & 0.35 & 0.34 & 0.05 & 0.04& --- & --- & --- & --- & --- \\
$  12.0 < {\rm R} < 13.0$ &   680 & 0.33 & 0.31 & 0.06 & 0.02& --- & --- & --- & --- & --- \\
$  13.0 < {\rm R} < 14.0$ &  1281 & 0.29 & 0.28 & 0.07 & 0.03& --- & --- & --- & --- & --- \\
$  14.0 < {\rm R} < 15.0$ &  2182 & 0.26 & 0.25 & 0.07 & 0.03 &   32 &  0.09& --0.09 &  0.04 & ---\\
$  15.0 < {\rm R} < 16.0$ &  3423 & 0.22 & 0.21 & 0.08 & 0.04 &  197 &  0.06& --0.05 &  0.04 &  0.03\\
$  16.0 < {\rm R} < 17.0$ &  5171 & 0.18 & 0.17 & 0.08 & 0.04 &  737 &  0.05& --0.02 &  0.04 &  0.03\\
$  17.0 < {\rm R} < 18.0$ &  7402 & 0.14 & 0.13 & 0.08 & 0.05 & 2556 &  0.05& --0.01 &  0.05 &  0.04\\
$  18.0 < {\rm R} < 19.0$ & 11035 & 0.11 & 0.09 & 0.09 & 0.06 & 7415 &  0.07&   0.00 &  0.07 &  0.07\\
$  19.0 < {\rm R} < 20.0$ & 19759 & 0.13 & 0.02 & 0.12 & 0.11 &11490 &  0.14&   0.02 &  0.14 &  0.13\\
$  20.0 < {\rm R} < 21.0$ & 26514 & 0.18 & 0.07 & 0.17 & 0.16 & 7061 &  0.26&   0.11 &  0.25 &  0.24\\
\end{tabular}
\caption[ ]{Internal systematic and random errors measured from the overlapping plate
pair AAO--R fields~292 and~241.}
\label{int_r}
\end{center}
\end{table*}
\begin{table*}
\begin{center}
\begin{tabular}{crccccrcccc}
 & \multicolumn{5}{c}{Stars} & \multicolumn{5}{c}{Galaxies} \\
Range & \multicolumn{1}{c}{N} & $\sigma_{\rm RAW}$ & ZP & $\sigma_{\rm -ZP}$ & $\sigma_{\rm REL}$  
& \multicolumn{1}{c}{N} & $\sigma_{\rm RAW}$ & ZP & $\sigma_{\rm -ZP}$ & $\sigma_{\rm REL}$ \\
\multicolumn{11}{c}{ }\\
$   9 < {\rm I} < 10$&    99 &  0.16& $  -0.14$&   0.10 &  ---& --- & --- & --- & --- & --- \\
$  10 < {\rm I} < 11$&   217 &  0.36& $  -0.34$&   0.08 &  ---& --- & --- & --- & --- & --- \\
$  11 < {\rm I} < 12$&   501 &  0.41& $  -0.39$&   0.07 &  0.04& --- & --- & --- & --- & --- \\
$  12 < {\rm I} < 13$&  1079 &  0.36& $  -0.35$&   0.08 &  0.04& --- & --- & --- & --- & --- \\
$  13 < {\rm I} < 14$&  1994 &  0.29& $  -0.28$&   0.07 &  0.04 &   13 &  0.29 &  0.27 &  0.10 &  ---\\
$  14 < {\rm I} < 15$&  3504 &  0.21& $  -0.20$&   0.07 &  0.04 &  132 &  0.19 &  0.17 &  0.13 &  ---\\
$  15 < {\rm I} < 16$&  5751 &  0.16& $  -0.15$&   0.07 &  0.05 &  530 &  0.14 &  0.12 &  0.09 &  0.06\\
$  16 < {\rm I} < 17$&  7827 &  0.11& $  -0.09$&   0.08 &  0.06 & 2006 &  0.11 &  0.07 &  0.09 &  0.08\\
$  17 < {\rm I} < 18$& 10050 &  0.11& $  -0.01$&   0.11 &  0.09 & 6195 &  0.14 & $-0.03$ &  0.14 &  0.14\\
$  18 < {\rm I} < 19$& 11213 &  0.17& $   0.01$&   0.17 &  0.16 & 5143 &  0.23 &  0.04 &  0.23 &  0.22\\
\end{tabular}
\caption[ ]{Internal systematic and random errors measured from the overlapping plate
pair SERC--I fields~292 and~241.}
\label{int_i}
\end{center}
\end{table*}
\begin{table*}
\begin{center}
\begin{tabular}{crccccrcccc}
 & \multicolumn{5}{c}{Stars} & \multicolumn{5}{c}{Galaxies} \\
Range & \multicolumn{1}{c}{N} & $\sigma_{\rm RAW}$ & ZP & $\sigma_{\rm -ZP}$ & $\sigma_{\rm REL}$  
& \multicolumn{1}{c}{N} & $\sigma_{\rm RAW}$ & ZP & $\sigma_{\rm -ZP}$ & $\sigma_{\rm REL}$ \\
\multicolumn{11}{c}{ }\\
$  9 < {\rm R} <  10$&    65 &  0.06 & $-0.01 $&  0.06   & ---& --- & --- & --- & --- & --- \\
$  10 < {\rm R} < 11$&   160 &  0.08& $ -0.07 $&  0.07&   0.05& --- & --- & --- & --- & --- \\
$  11 < {\rm R} < 12$&   340 &  0.16& $ -0.15 $&  0.07&   0.06& --- & --- & --- & --- & --- \\
$  12 < {\rm R} < 13$&   621 &  0.22& $ -0.21 $&  0.08&   0.03& --- & --- & --- & --- & --- \\
$  13 < {\rm R} < 14$&   983 &  0.25& $ -0.24 $&  0.09&   0.04& --- & --- & --- & --- & --- \\
$  14 < {\rm R} < 15$&  1461 &  0.23& $ -0.22 $&  0.10&   0.04 & 35 & 0.05& $  -0.02 $&  0.05  & 0.05\\
$  15 < {\rm R} < 16$&  2276 &  0.18& $ -0.17 $&  0.10&   0.05 & 174 &  0.12& $  -0.10 $&  0.09&   0.07\\
$  16 < {\rm R} < 17$&  3630 &  0.14& $ -0.13 $&  0.10&   0.05 & 502 &  0.09& $  -0.06 $&  0.08&   0.04\\
$  17 < {\rm R} < 18$&  5041 &  0.13& $ -0.11 $&  0.10&   0.06 & 1384&  0.09& $  -0.05 $&  0.09&   0.06\\
$  18 < {\rm R} < 19$&  6225 &  0.13& $ -0.09 $&  0.12&   0.08 & 3766&  0.15& $  -0.08 $&  0.15&   0.11\\
$  19 < {\rm R} < 20$&  8576 &  0.19& $ -0.08 $&  0.18&   0.15 &5445 &  0.22& $  -0.08 $&  0.22&   0.19\\
$  20 < {\rm R} < 21$&  4408 &  0.19& $  0.09 $&  0.18&   0.17 &1379 &  0.30& $   0.24 $&  0.25&   0.22\\
\end{tabular}
\caption[ ]{Internal systematic and random errors measured from the overlapping plate
pair ESO--R fields~292 and~241.}
\label{int_s}
\end{center}
\end{table*}

The following points should be noted from Tables~\ref{int_j} to~\ref{int_s}:
\begin{itemize}
\item There are no systematic zeropoints at the faint end -- this is of course
a result of the plate matching procedure detailed in Section~\ref{global}.
\item There are increasingly large systematic errors for increasingly bright
stellar magnitude ranges.
\item Galaxy magnitudes are less susceptible to systematic effects than are
stellar magnitudes.
\item The {\em relative} precision of stellar magnitudes (i.e.~within 
restricted position and magnitude ranges) is 5\% or better when 3~mag or more
above the respective plate limits.
\item Relative galaxy magnitudes are, in general, worse than those of stars in 
a given magnitude range, as expected.
\end{itemize}
The systematic errors for bright stellar images are partly due to positional
effects and partly due to the assumption of a mean gradient for the linear
part of the profile magnitude scale. Since the gradient of the isophotal
scale is also tied to this mean gradient, then systematic effects are also
increasingly apparent in increasingly bright galaxian magnitude ranges due to
the linear extrapolation from the faint levels at which the zeropoint is fixed.
Section~\ref{stone} shows some of these effects when comparing SGC photometry
against an independent, external dataset.

\begin{figure*}
\begin{center}
\setlength{\unitlength}{1mm}
\begin{picture}(150,220)
\includegraphics{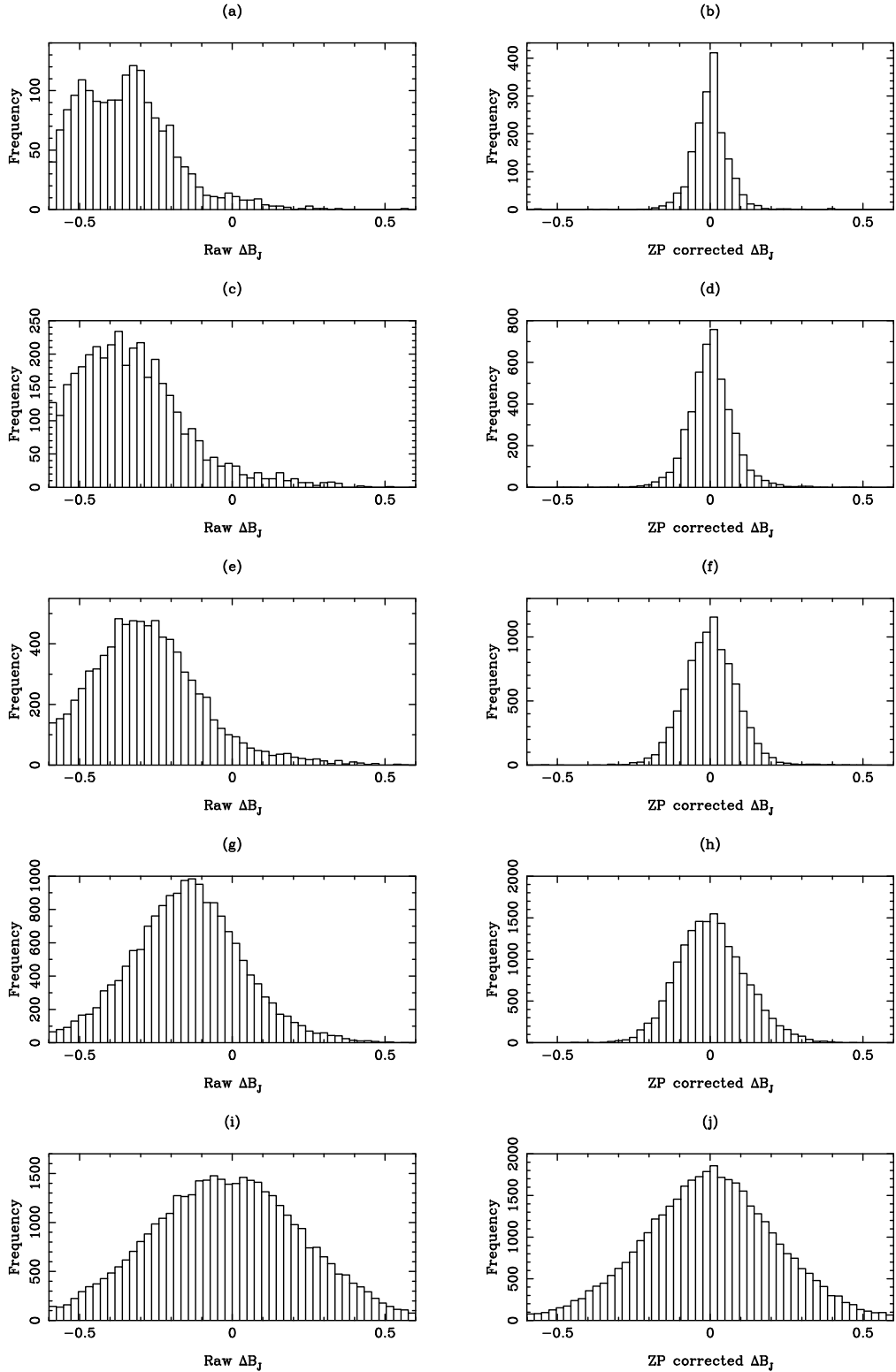}
\end{picture}
\end{center}
\caption[]{Histograms of differences between J~survey photometry measured
in the overlap region of fields~292 and~241 for the magnitude ranges
(a,b) $13<{\rm B}_{\rm J}<15$; 
(c,d) $15<{\rm B}_{\rm J}<17$; 
(e,f) $17<{\rm B}_{\rm J}<19$; 
(g,h) $19<{\rm B}_{\rm J}<21$; and
(i,j) $21<{\rm B}_{\rm J}<23$. The left--hand panels are raw differences; the 
right--hand panels show distributions after subtraction of positionally
dependent systematic errors.}
\label{jrelphothist}
\end{figure*}

Figure~\ref{jrelphothist}(a) to~(j) illustrates some of these effects graphically,
where we have plotted stellar J~survey data in five magnitude ranges of 2~mag
size in the range $13<{\rm J}<23$. The left--hand panels show histograms of raw
magnitude differences, while the right--hand panels show the underlying
distributions of random errors left over after positionally--dependent  
systematic errors have been removed. Once again, the increase in random error,
and decrease in systematic error, with decreasing brightness are apparent.

\subsubsection{Photometry (colours)}

\begin{figure*}
\begin{center}
\setlength{\unitlength}{1mm}
\begin{picture}(150,50)
\includegraphics{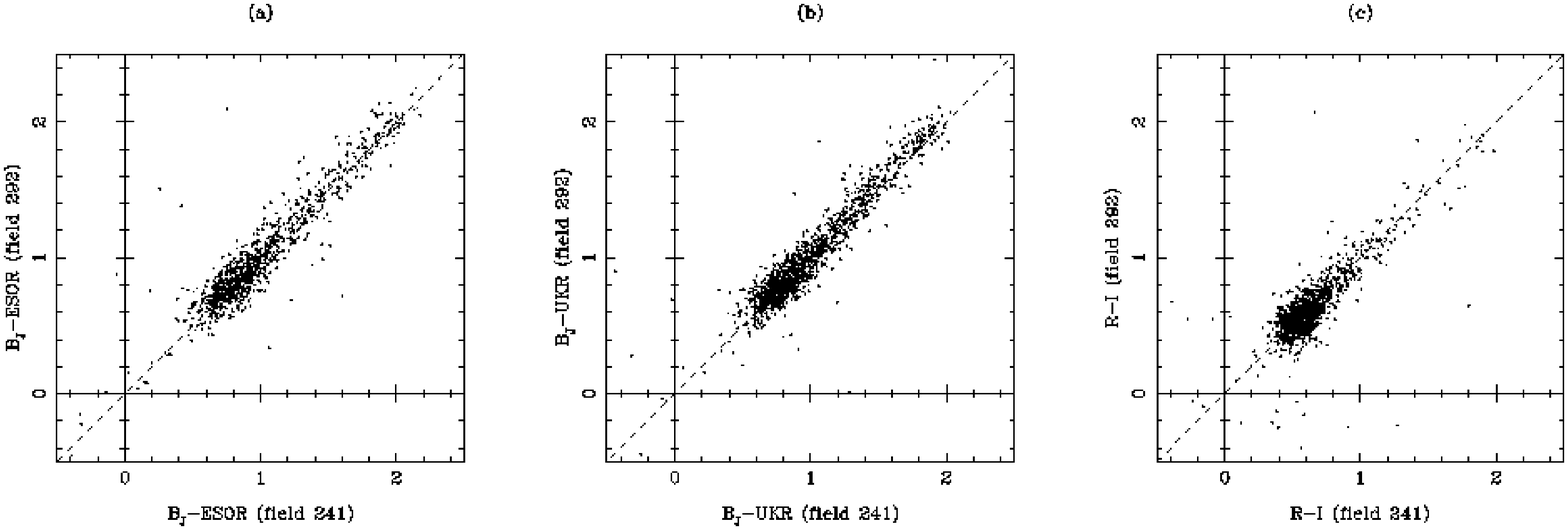}
\end{picture}
\end{center}
\caption[]{Comparisons of survey colours in the overlap region of fields~241
and~292 in the magnitude range $16<{\rm B}_{\rm J}<17$:
(a) B$_{\rm J}-R$ (ESO--R); (b)  B$_{\rm J}-R$ (UKR) and (c) R--I.
The dashed lines shows $y=x$. The RMS dispersion is $\sim0.1$~mag indicating
single--measurement errors of $\sim0.07$~mag in any colour in this
magnitude range.}
\label{colover}
\end{figure*}

Figure~\ref{colover} shows the consistency of our survey colours J--R (with~R
from both UK and ESO Schmidts) and R--I in the magnitude range 
$16<{\rm B}_{\rm J}<17$. Once again, because we have calibrated out 
the systematic
errors in colour as a function of position and magnitude, and also set the 
colour zeropoint in every field (see Section~\ref{colbias}) the colours are
consistent at a level dictated by random errors alone. The RMS values about the
$y=x$ lines in Figure~\ref{colover} are $\sim0.1$~mag, indicating 
single measurement random errors in the colours at the level of $\sim0.07$~mag
in this magnitude range. While this is strictly
speaking a test of internal consistency, it also illustrates the true external
errors in the colours (cf.~Section~\ref{colext}) owing to the way in which any
systematic errors have been eliminated.

\subsection{External tests of accuracy and reliability}

\subsubsection{Image classification and completeness}
\label{extsg}

We have assessed the accuracy of image classification and detection completeness
of SGC data using 4m prime focus data from Carter~(1980) and CCD data
from Metcalfe et al.~(1991 -- hereafter MSFJ91). Those data are demonstrably 
deeper than that from the Schmidt plates and although they are only slightly
better in terms of angular resolution (e.g.~1.5~arcsec for the MSFJ91 CCD images
as opposed to $\sim2$~arcsec for the SGC data) their higher signal--to--noise
ratio at a given magnitude allows thresholding at lower isophotes and therefore
more reliable image classification. In Table~\ref{carter} we give figures for
comparison against the galaxy sample of Carter~(1980). This comparison shows
completeness levels of 100\% down to B$_{\rm J}\sim21$ with $>90$\% reliability
of correct classification down to B$_{\rm J}\sim20$. This is in good
agreement with other estimates of the reliability of Schmidt survey galaxy 
catalogues (e.g.~Caretta, Maia \& Willmer~2000). 
\begin{table*}
\begin{center}
\begin{tabular}{crrrrrr}
Magnitude range & No.~of gals in & No.~of objects & No.~classed & No.~classed &
 Completeness & Reliabilty \\
                & Carter (1980)  & found in SGC   & as galaxies & as stellar &
           \%   &     \%       \\
\multicolumn{7}{c}{ } \\
$ 16.5 < {\rm B}_{\rm J} < 17.5$ &   3 &   3 &   3 &   0  &   100.0  &   100.0 \\
$ 17.5 < {\rm B}_{\rm J} < 18.5$ &  28 &  28 &  28 &   0  &   100.0  &   100.0 \\
$ 18.5 < {\rm B}_{\rm J} < 19.5$ &  92 &  91 &  91 &   0  &    98.9  &   100.0 \\
$ 19.5 < {\rm B}_{\rm J} < 20.5$ & 180 & 174 & 160 &  14  &    96.7  &    92.0 \\
$ 20.5 < {\rm B}_{\rm J} < 21.5$ & 307 & 302 & 227 &  75  &    98.4  &    75.2 \\
$ 21.5 < {\rm B}_{\rm J} < 22.5$ & 584 & 515 & 253 & 262  &    88.2  &    49.1 \\
$ 22.5 < {\rm B}_{\rm J} < 23.5$ & 350 & 149 &  57 &  92  &    42.6  &    38.3 \\
\end{tabular}
\caption[ ]{Comparison of the galaxy sample of Carter~(1980) against the
corresponding SGC J data.}
\label{carter}
\end{center}
\end{table*}
Tables~\ref{eqj} and~\ref{eqr} give similar figures for comparison of sets
of stars and galaxies from MSFJ91 against SGC~J and~R data. Field numbers~823
to~826 and 891, 893 and~894 were used in these comparisons (fields 4, 13, 
10, 14, 2, 3 and~9 respectively in Table~1 of MSFJ91).
\begin{table*}
\begin{center}
\begin{tabular}{crrrrrrrrrrrr}
                & \multicolumn{6}{c}{Stars} & \multicolumn{6}{c}{Galaxies} \\
Magnitude range & Tot. & SGC & Stars & Gals & \%Comp. & \%Rel. &
 Tot. & SGC & Gals & Stars & \%Comp. & \%Rel. \\
\multicolumn{13}{c}{ }\\
$ 16.5 < {\rm B}_{\rm J} < 17.5$&    8&    8&    8&    0&  100&  100& --- & --- & --- & --- & --- & --- \\
$ 17.5 < {\rm B}_{\rm J} < 18.5$&    7&    7&    7&    0&  100&  100&    3&    3&    3&    0&  100&  100\\
$ 18.5 < {\rm B}_{\rm J} < 19.5$&   12&   11&   11&    0&   92&  100&    4&    4&    4&    0&  100&  100\\
$ 19.5 < {\rm B}_{\rm J} < 20.5$&   20&   19&   19&    0&   95&  100&   14&   14&   12&    2&  100&   86\\
$ 20.5 < {\rm B}_{\rm J} < 21.5$&   42&   22&   18&    4&   52&   82&   53&   53&   33&   20&  100&   62\\
$ 21.5 < {\rm B}_{\rm J} < 22.5$&  117&   34&   31&    3&   29&   91&  129&   76&   25&   51&   59&   33\\
$ 22.5 < {\rm B}_{\rm J} < 23.5$&  174&    2&    2&    0&    1&  100&  363&   17&    7&   10&    5&   41\\
\end{tabular}
\caption[ ]{Comparison of the star/galaxy samples from Metcalfe et al.~(1991) against the corresponding SGC~J data.}
\label{eqj}
\end{center}
\end{table*}
\begin{table*}
\begin{center}
\begin{tabular}{crrrrrrrrrrrr}
                & \multicolumn{6}{c}{Stars} & \multicolumn{6}{c}{Galaxies} \\
Magnitude range & Tot. & SGC & Stars & Gals & \%Comp. & \%Rel. &
 Tot. & SGC & Gals & Stars & \%Comp. & \%Rel. \\
\multicolumn{13}{c}{ }\\
$ 16.5 < {\rm R} < 17.5$&    3&    3&    3&    0&  100&  100&    8&    8&    7&    1&  100&   88\\
$ 17.5 < {\rm R} < 18.5$&   11&   11&   10&    1&  100&   91&   20&   20&   18&    2&  100&   90\\
$ 18.5 < {\rm R} < 19.5$&   31&   31&   12&   19&  100&   39&   25&   24&   21&    3&   96&   88\\
$ 19.5 < {\rm R} < 20.5$&   94&   80&   27&   53&   85&   34&   25&   23&   18&    5&   92&   78\\
$ 20.5 < {\rm R} < 21.5$&  189&   26&   11&   15&   14&   42&   31&    6&    4&    2&   19&   67\\
\end{tabular}
\caption[ ]{Comparison of the star/galaxy samples from Metcalfe et al.~(1991) against the SGC~R
data.}
\label{eqr}
\end{center}
\end{table*}

The reliability figures in Tables~\ref{carter} and~\ref{eqj} clearly do not
agree with the internally assessed classification accuracy estimates displayed
in Figures~\ref{clsprob}(a) and~(b). The reason for this is that fainter than
m~$\sim20$, the typical size of detectable galaxian images is comparable to the
seeing on the photographic plate material, and the image classification scheme
inevitably classifies unresolved sources as stars. In Figures~\ref{clsprob}(a)
to~(b) the accuracy of the classification is overestimated since there is a
systematic trend for objects to be classed as stellar regardless of their
actual type. The false level of agreement on stellar classification potentially 
leads to a
correspondingly false estimate of the underlying classification accuracy for
both image types; however it is clear from the internal reliability that some
very faint, non--stellar sources must be consistenly classified as such when it
is possible to detect their non--stellar nature from the photographic data.

It is interesting to note that the reliability estimates from Table~\ref{carter}
were made from a field which presents a small zenith distance at the UK Schmidt
Telescope, while Tables~\ref{eqj} and~\ref{eqr} are based on data at the 
celestial equator. This probably accounts for their poorer image classification
reliability.

\subsubsection{Systematic photometric errors as a function of magnitude}
\label{stone}

\begin{figure}
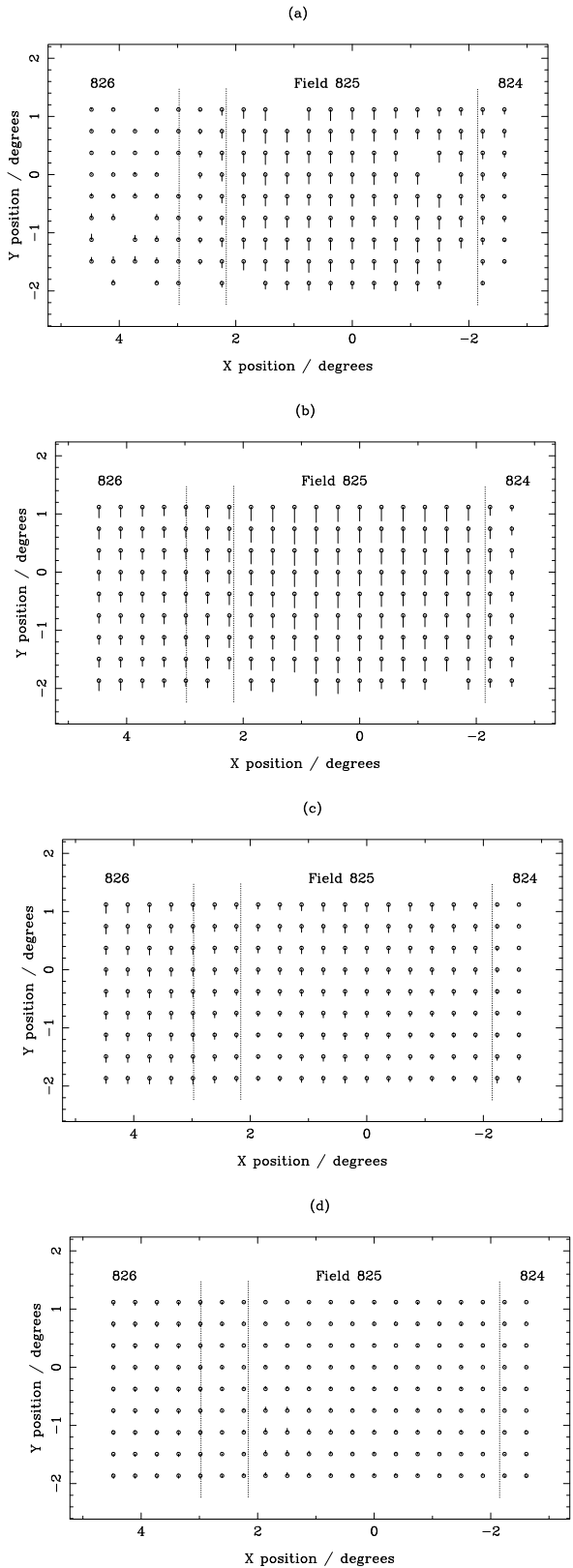

\begin{center}
\setlength{\unitlength}{1mm}
\begin{picture}(80,210)
\includegraphics{figure15a.eps}
\includegraphics{figure15b.eps}
\includegraphics{figure15c.eps}
\includegraphics{figure15d.eps}
\end{picture}
\end{center}
\caption[]{Positional dependent systematic errors between SGC R data and the
CCD photometry of Stone et al.~(1999) in SDSS calibration region `A'. Data
are binned in $\sim2$~cm bins and smoothed to show the global trends in the
systematics; in plots~(a) to~(d) we show four 2.0~mag 
ranges between $10<{\rm R}<18$. The vertical bars are scaled such that
$1^{\circ}\equiv1$ mag.}
\label{stonemag}
\end{figure}

As illustrated in Figure~\ref{jrelphothist}, while the survey field
photometric zeropoints are defined
globally using faint objects in overlap regions along with a network
of CCD standards, at increasingly bright magnitudes systematic errors become
apparent. This is due to the assumption of a single global gradient for the
characteristic curve for the plates in any given colour. Furthermore, it is to be
expected that systematic errors will be dependent on plate position as well as
magnitude, since the photographic response will not be perfectly uniform within
a single plate. This is illustrated in more detail in Figure~\ref{stonemag}(a)
to~(d), where we have compared SGC R photometry against the SDSS calibration
region~A (Stone, Pier \& Monet~1999). These data are in a slightly different
passband (SLOAN
r as oppose to R$_{\rm 59F}$) but the difference is negligible at the
accuracy we require, and the random errors on the CCD data are negligible 
compared to the photographic errors. The CCD data encompass 
$\sim3.5^{\circ}\times7.5^{\circ}$ and are therefore ideal for showing position
dependent systematics over areas that traverse photographic plate boundaries.

In Figure~\ref{stonemag}(a) to~(d) we have divided the data in $10<{\rm R}<18$
into four 2~mag ranges; the vertical error vectors are scaled such that 
$1^{\rm m}\equiv1^{\circ}$. Only stars have been used in this analysis. The
following points should be noted:
\begin{itemize}
\item In the faintest magnitude range (${\rm R}>16$) there are no significant
zeropoint errors within a field or in traversing to an adjacent field.
\item There are global systematic zeropoint errors within individual fields at
the level of several tenths of a magnitude for ${\rm R}<16$; these are, in
general, different for different fields.
\item At bright magnitudes, the systematic errors change within a single field
as a function of plate position. These variations are once more at the level
of several tenths of a magnitude.
\end{itemize}

\subsubsection{Effective external errors in photometry}
\label{colext}

Finally, we have checked the external errors (i.e.~those resulting in a
straightforward comparison between an external dataset and our survey data,
with no regard as to the systematic nature of those errors) of our BRI
photometry in the equatorial zone by combining data from Landolt~(1992) and
Boyle et al.~(1995). Note that although the Boyle et al.\ data have been used
to measure gradients and tie down zeropoints (Sections~\ref{grads} 
and~\ref{global}), they have not been used to derive detailed individual
calibration curves on a plate--by--plate basis. Galaxy photometric precision
has been assessed using the Metcalfe et al.~(1991) data along with data from
Cunow et al.~(1997 and references therein). Table~\ref{absacc} shows the
results. The main point to note from this comparison is that brighter than
m~$\sim15$, there can be large errors in the photometry (their origin is
discussed previously). Fainter than this, in general the stellar photometry
is typically accurate to $\sigma\sim0.3$ (at best, $\sigma\sim0.1$) until the
respective plate limits are approached, whereupon emulsion noise begins to
dominate and the random errors rise again (this is particularly noticeable for
the lower s/n SERC--I and POSS--E plates). It is also clear that galaxy 
photometry is, in general, less accurate than stellar photometry in a given
magnitude range as expected.

\begin{figure}
\begin{center}
\setlength{\unitlength}{1mm}
\begin{picture}(80,40)
\includegraphics{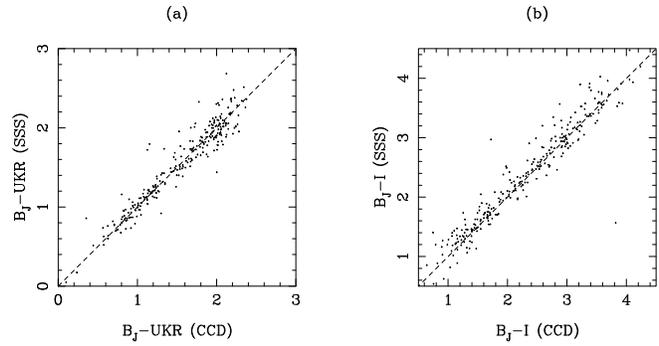}
\end{picture}
\end{center}
\caption[]{Comparisons of survey colours with those from Boyle et al.~(1995):
(a) B$_{\rm J}-{\rm R}$; (b)  B$_{\rm J}-{\rm I}$.
The dashed lines shows $y=x$. The RMS dispersion is $\sim0.16$~mag indicating
the single--measurement errors in any colour at these magnitudes
(typically B$_{\rm J}\sim20$).}
\label{extcol}
\end{figure}

Figure~\ref{extcol} illustrates external checks on the colour indices
B$_{\rm J}$--R and B$_{\rm J}$--I from the same
data, in the natural systems of the photographic emulsion/filter passbands.
These indicate the external accuracy of the SSS colours. Despite the presence
of large systematic errors on individual magnitudes, the colours are accurate
to the levels indicated in the previous internal consistency check: at
B$_{\rm J}\sim16.5$, the colours are accurate to the level 
$\sigma_{\rm B-R,R-I}\sim0.07$; at B$_{\rm J}\sim20.0$ 
(typical of the external comparison stars in Figure~\ref{extcol}) the errors
rise to $\sigma_{\rm B-R,R-I}\sim0.16$. Moreover, there is no evidence of
any systematic errors in the colours greater than $\sim0.1$~mag, vindicating the
choice of colour zeropoints given in Table~\ref{blueshifts}.

\begin{table*}
\begin{center}
\begin{tabular}{crrrrrrrrrrrrrrrr}
                & \multicolumn{4}{c}{SERC--EJ} & \multicolumn{4}{c}{SERC--ER} 
                & \multicolumn{4}{c}{SERC--I} & \multicolumn{4}{c}{POSS--E} \\
 & N$_{\rm S}$ & $\sigma$ & N$_{\rm G}$ & $\sigma$ 
 & N$_{\rm S}$ & $\sigma$ & N$_{\rm G}$ & $\sigma$ 
 & N$_{\rm S}$ & $\sigma$ & N$_{\rm G}$ & $\sigma$ 
 & N$_{\rm S}$ & $\sigma$ & N$_{\rm G}$ & $\sigma$ \\
\multicolumn{17}{c}{ }\\
$10<{\rm m}<11$ &  ---  &  ---  & ---  & ---   &  ---  & ---   & ---   & --- &     5 &  0.97 &  --- &  --- &     4 &  0.77 &  --- &  --- \\
$11<{\rm m}<12$ &   --- & ---   &  ---  & ---   &   5 &  0.66 &  --- &  --- &    11 &  0.37 &  --- &  --- &     17 &  0.96 &  --- &  --- \\
$12<{\rm m}<13$ &   --- &  ---  & ---   & ---   &   7 &  0.69 &  --- &  --- &      7 &  1.01 &  --- &  --- &     9 &  1.13 &  --- &  --- \\
$13<{\rm m}<14$ &     5 &  0.33 &   --- &  --- &   4 &  0.66 &  --- &  --- &     9 &  0.93 &  --- &  --- &     7 &  1.11 &  --- &  --- \\
$14<{\rm m}<15$ &     5 &  0.38 &   --- &  --- &   6 &  0.54 &  --- &  --- &     7 &  0.72 &  --- &  --- &    11 &  0.85 &  --- &  --- \\
$15<{\rm m}<16$ &    10 &  0.33 &  --- &  --- &   6 &  0.40 &  --- &  --- &    27 &  0.21 &   --- &  --- &     7 &  0.73 &   --- &  --- \\
$16<{\rm m}<17$ &    25 &  0.22 &   --- &  --- &  38 &  0.15 &     5 &  0.73 &    44 &  0.10 &     6 &  0.29 &    62 &  0.16 &     6 &  0.76 \\
$17<{\rm m}<18$ &    47 &  0.14 &   --- &  --- &  64 &  0.08 &    19 &  0.16 &    70 &  0.09 &    39 &  0.19 &    92 &  0.17 &    25 &  0.27 \\
$18<{\rm m}<19$ &    75 &  0.10 &    19 &  0.85 &  96 &  0.07 &    56 &  0.16 &    74 &  0.17 &    46 &  0.37 &   148 &  0.22 &    78 &  0.24 \\
$19<{\rm m}<20$ &    90 &  0.09 &    46 &  0.37 & 129 &  0.12 &    61 &  0.14 &    15 &  0.84 &     7 &  0.62 &   162 &  0.29 &    91 &  0.35 \\
$20<{\rm m}<21$ &   160 &  0.11 &    85 &  0.25 &  50 &  0.25 &    15 &  0.31 &       ---&   --- &  ---  & ---   &     39 &  1.31 &    20 &  1.40 \\
$21<{\rm m}<22$ &   239 &  0.23 &    51 &  0.21 &  ---   &  ---  &  ---  &  ---  &        ---&  ---  &  ---  &   --- &    ---  &  ---  &  ---  &  ---  \\
$22<{\rm m}<23$ &    27 &  0.22 &     3 &  0.22 &  ---   &  ---  &  ---  &  ---  &        ---&  ---  &  ---  &   --- &    ---  &  ---  &  ---  &  ---  \\
\end{tabular}
\caption[ ]{Global external photometric
accuracy of SGC data as a function of magnitude in
the equatorial region for stars and galaxies (N$_{\rm S}$ and N$_{\rm G}$
respectively).}
\label{absacc}
\end{center}
\end{table*}

\section{Conclusion}
\label{concs}

We have presented a detailed description of the image detection, 
parameterisation, classification and photometric properties of the
SuperCOSMOS Sky Survey\footnote{database available
online at \verb+http://www-wfau.roe.ac.uk/sss+}. 
With reference to examples from the first release
of data, the South Galactic Cap (SGC) survey, we have demonstrated
the typical completeness, classification reliability and photometric errors
(both random and systematic). For a summary of
these results along with similar comparisons for other large--scale survey
programmes and a guide to using the SSS database, the reader is
referred to Paper~{\sc I}.

\section*{Acknowledgements}

The SuperCOSMOS pixel analysis software is based on original code written by
Steven Beard for the predecessor machine, COSMOS. We are also grateful to
Dennis Kelly for the huge coding effort in writing new routines and importing
existing software into the SuperCOSMOS suite. NCH is grateful to Simon Driver
for discussion concerning galaxy magnitudes, to Lance Miller for his computer 
codes for colour corrections and to Sue Tritton and Mike Read
for information concerning the various Schmidt survey plate collections.
We thank Nigel Metcalfe and David Carter
for supplying object catalogues from deep
CCD and 4m photographic data in machine--readable form.
Funding for the University of Edinburgh Institute
for Astronomy Wide--Field Astronomy Unit and Institute of Astronomy Cambridge
Astronomical Survey Unit is provided by the UK PPARC. This research has made
use of data archived at the CDS, Strasbourg. 
We are indebted to the referee, Sean Urban, for a prompt and 
thorough review of these manuscripts.

The National Geographic Society--Palomar Observatory Sky Survey (POSS--I) was
made by the California Institute of Technology with grants from the National
Geographic Society. The UK Schmidt Telescope was operated by the Royal
Observatory Edinburgh, with funding from the UK Science and Engineering 
Research Council (later the UK Particle Physics and Astronomy Research Council),
until 1988~June, and thereafter by the Anglo--Australian Observatory. The blue
plates of the Southern Sky Atlas and its equatorial extension (together known
as the SERC--J/EJ) as well as the Equatorial Red (ER), the second epoch
(red) Survey (SES or AAO--R) and the infrared (SERC--I) Survey
were taken with the UK Schmidt Telescope. All data
retrieved from the URLs described herein
are subject to the copyright given in this copyright summary. Copyright
information specific to individual plates is provided in the downloaded FITS
headers.

\vfill
\bsp

\end{document}